\documentclass[useAMS,usenatbib]{mn2e}

\usepackage{latexsym,amsmath,amstext} 
\usepackage[dvips]{graphicx}          

\usepackage{txfonts}
\usepackage{graphicx}
\usepackage{natbib}
\bibpunct{(}{)}{;}{a}{}{,}

\title[Kozai problem with an eccentric perturber]
{Interesting dynamics at high mutual inclination in the framework of the Kozai problem with an eccentric perturber}

\author[A.-S. Libert and N. Delsate]{A.-S. Libert$^{1}$\thanks{E-mail:
anne-sophie.libert@fundp.ac.be (ASL); nicolas.delsate@fundp.ac.be (ND)} and N. Delsate$^{1}$\\
$^{1}$ NaXys, Department of Mathematics FUNDP, 8 Rempart de la Vierge, B-5000 Namur, Belgium}

\begin{document}

\pagerange{\pageref{firstpage}--\pageref{lastpage}} \pubyear{2009}

\maketitle

\label{firstpage}

%
%
%

\begin{abstract} 
We study the dynamics of the 3-D three-body problem of a small body moving under the attractions of a star and a giant planet which orbits the star on a much wider and elliptic orbit. In particular, we focus on the influence of an eccentric orbit of the outer perturber on the dynamics of a small highly inclined inner body. Our analytical study of the secular perturbations relies on the classical octupole hamiltonian expansion (third-order theory in the ratio of the semi-major axes), as third-order terms are needed to consider the secular variations of the outer perturber and potential secular resonances between the arguments of the pericenter and/or longitudes of the node of both bodies. Short-period averaging and node reduction (by adoption of the Laplace plane reference frame) reduce the problem to two degrees of freedom. The four-dimensional dynamics is analyzed through representative planes which identify the main equilibria of the problem. As in the circular problem (i.e. perturber on a circular orbit), the ``Kozai-bifurcated" equilibria play a major role in the dynamics of an inner body on quasi-circular orbit: its eccentricity variations are very limited for mutual inclination between the orbital planes smaller than $\sim 40^\circ$, while they become large and chaotic for higher mutual inclination.  Particular attention is also given to a region around $35^\circ$ of mutual inclination, detected numerically by \citet{Fun11} and consisting of long-time stable and particularly low eccentric orbits of the small body.  Using a 12th-order Hamiltonian expansion in eccentricities and inclinations, in particular its action-angle formulation obtained by Lie transforms in \citet{Lib08}, we show that this region presents an equality of two fundamental frequencies and can be regarded as a secular resonance. Our results also apply to binary star systems where a planet is revolving around one of the two stars.
\end{abstract}



\section{Introduction}

The (inner) {\it Lidov-Kozai mechanism} (Kozai 1962; Lidov 1962) is a well-known secular resonance of the restricted three-body problem, which can be reduced two degrees of freedom after short-period averaging and node reduction (see for instance Malige et al. 2002). Kozai (1962) showed that a highly inclined asteroid perturbated by Jupiter periodically exchanges its eccentricity and inclination. Its analytical theory relied on the assumption that Jupiter's orbit is circular, so that the problem is integrable. Since its discovery, the Kozai resonance has found numerous applications in studies of planetary and stellar systems.

Recently, analytical studies (e.g. Michtchenko et al. 2006, Libert \& Henrard 2007) have shown the possibility that extrasolar planetary systems can be in a long-term stable highly non-coplanar configuration, sometimes due to a secular Kozai-type phase-protection mechanism. For instance, Libert \& Tsiganis (2009) found that $\upsilon$ Andromedae, HD 12661, HD 169830 and HD 74156 extrasolar two-planet systems have orbital parameters compatible with a Kozai-resonant state, if their (unknown) mutual inclination is at least $45^\circ$. 

The Kozai dynamical phenomenon is also well-known in studies of binary systems (e.g. Innanen et al. 1997, Wu \& Murray 2003, Fabrycky \& Tremaine 2007), in particular in the S-type configuration (a planet revolves around one of the primaries) where the orbit of a highly inclined planet can undergo large amplitude oscillations of its eccentricity. A similar Kozai-resonant evolution can be observed in the C-type configuration, where the planet is in orbit around the binary (e.g. \citealt{Mig11}). 

Concerning the planetary three-body problem, the discovery of giant extrasolar planets on eccentric orbit raises the question of the influence of their eccentricity on potential asteroids and Earth-mass companions on inclined orbit. In a preliminary numerical study of Funk et al. (2011), the long-term stability of inclined fictitious Earth-mass planets in the habitable zone of extrasolar giant planets discovered so far is analyzed. They have realized a parametric analysis, where several giant planet's eccentricities are considered, while the Earth-like body is initially on a circular orbit, closer to the star than the gas giant. Their simulations show that, for distant orbits, test-planets below a critical inclination of approximately $40^\circ$ are in a stable configuration with gas giants on either circular (i.e. well-known result associated with the Kozai mechanism) or elliptic orbit. Furthermore, for gas giant on eccentric orbit, the small companion exhibits non-negligeable variations in eccentricity, except for a region around $35^\circ$ of mutual inclination of the orbital planes, consisting of long-time stable and low eccentric orbits of the Earth-like body. So the influence of the eccentricity of the perturber on the dynamics of inclined Earth-mass planets seems to be significant and deserve to be studied in more detail with dynamical tools. This is the goal of the present contribution. 

In this work, we focus on the 3-D three-body problem of a small body moving under the attractions of a star and a giant planet which orbits the star on a much wider and elliptic orbit. Our analytical study of the secular perturbations relies on the classical octupole hamiltonian expansion (third-order theory in the ratio of the semi-major axes), widely used in planetary and stellar systems (e.g. Ford et al. 2000, Lee \& Peale 2003, Migaszewski \& Go\'zdziewski 2011). Actually, third-order terms are needed to introduce the secular variations of the eccentricity of the perturber. Indeed, the second-order quadrupole approximation does not depend on the argument of the pericenter of the perturber, whose eccentricity is thus an integral of motion (e.g. \citealt{Har69}, \citealt{Lid76}, \citealt{Fer94} and \citealt{Far10}). The third-order terms introduce qualitative changes in the dynamics and can explain the aformentioned dynamical features observed by Funk et al. (2011), as we will show in this work. Let us note that, even if the octupole development is an analytical expansion of the three-body problem whatever their masses, we only focus on planetary systems with a small value of the inner body's mass hereafter. This problem is sometimes called the {\it reduced} spatial three-body problem. Since we consider the (very limited) effect of the small mass on its companion, the secular evolution of the outer body is considered, and so are the potential secular resonances between the arguments of the pericenter and/or longitudes of the node of both bodies.

The paper is organized as follows. In Section 2, the octupole Hamiltonian formulation is recalled. Section 3 analyzes the four-dimensional secular dynamics of the elliptic spatial three-body problem, by means of 2-D geometric representations called {\it representative planes}. Section 4 focusses on the dynamical feature around $35^\circ$ of mutual inclination of the orbital planes, described in Funk et al. (2011). Finally our results are summarized in Section 5.

\section{Octupole Hamiltonian formulation}
Let us consider a system consisting of an inner small body ($m_1$) and an outer giant planet ($m_2$) orbiting a star ($m_0$) (also called inner three-body problem, see \citealt{Far10}). Due to their masses, the inner body will be named the {\it perturbed body} and the outer one the {\it perturber} in the following. We focus on the {\it spatial} (or 3-D) problem where both planetary orbits are mutually inclined. In the Solar System, this configuration corresponds for instance to an asteroid perturbated by Jupiter. Let us note that studies on the secular evolution of asteroids are mostly realized under the assumption that Jupiter's orbit is circular (e.g. \citealt{Koz62}). Since the discovery of extrasolar systems, inclined test particles, representing Earth-mass planets with weak gravitational effects on a system composed of a star and a gas giant, are another application of the spatial problem. However, as many giant extrasolar planets have eccentric orbit, one may wonder the influence of the eccentricity of such a Jupiter-like planet on its Earth-mass companion(s). To address this question, we consider in the following that perturber is on an eccentric orbit.

The spatial model of the three-body problem can be described using the {\it canonical heliocentric formulation} (see \citealt{Poi96}, \citealt{Las95}):
\begin{equation}  \label{democratic}
{\mathcal H}= \sum_{j=1}^2 \left\{\frac{ {\bf p}_j^2}{2m'_j} -  \frac{G(m_0+m_j)m'_j}{r_j} \right\}- G \frac{m_1m_2}{\|{\bf r}_1-{\bf r}_2\|}+\frac{{\bf p}_1 \cdot {\bf p}_2}{m_0} \, .
\end{equation}
where ${\bf r}_i$ are the position vectors of $m_i$ relative to the star, ${\bf p}_i$ are their conjugate momenta relative to the barycenter of the three-body system, and $m'_j=(1/m_0+1/m_j)^{-1}$ are the reduced masses. Let us recall that the heliocentric velocities ${\bf \dot{r}}_i$ will not follow the direction given by ${\bf p}_i$, and thus the ellipses are not tangent to the real trajectory. The first term of the expansion is the sum of the keplerian motions of the two planets. The perturbation of this integrable part, representing the mutual interactions between the planets, consists of the direct part and the indirect part respectively.

A set of canonical variables is formed by use of the classical Delaunay's elements:
\begin{equation}
\begin{tabular}{ll}
\hspace{-2cm} $l_j= M_j$, & $L_j=m'_j \sqrt{G(m_0+m_j)a_j}, $ \\ 
\hspace{-2cm}$g_j=\omega_j$, &  $G_j=L_j \sqrt{1-{e_j}^2}, $\\
\hspace{-2cm}$h_j=\Omega_j$, &  $H_{j}=G_{j} \cos{i_{j}}, $ 
\end{tabular}
\end{equation}
where $a_j$ denote semi-major axes of the planets, $e_j$ eccentricities, $i_j$ inclinations, $\omega_j$ arguments of the pericenter, $\Omega_j$ longitudes of ascending nodes, and $M_j$ mean anomalies, all being canonical heliocentric elements.

As we are interested in the long-term dynamics and assuming that we are not close to a mean motion resonance, we can average (to first order in the mass ratio) the Hamiltonian function over the fast angles, namely the mean anomalies $M_i$ (Deprit 1969). It means that the averaged Hamiltonian ${\mathcal K}$ does not depend on the mean anomalies, then the conjugate momenta $L_i$ are constants in the secular problem and so are the semi-major axes. So it results in a four degree of freedom formulation of the Hamiltonian function.

To average the indirect part of the disturbing function, we compute $\frac{1}{(2\pi)^2}\int_0^{2\pi}\int_0^{2\pi} {\bf \dot{p}}_i \cdot {\bf \dot{p}}_j \;\;dM_i dM_j=\delta_{ij}a_i^2 n_i^2$, where ${\bf p}_i$, ${\bf p}_j$ are canonical heliocentric velocities related to the canonical heliocentric elements.

Concerning the direct part, we use the traditional expansion in Legendre polynomials, assuming that $r_1<<r_2$:
\begin{equation}\label{Pn}
  H_{DP} =  -G\,m_1\,m_2 \frac{1}{r_2} \sum_{n\geq 0}^{\infty} \left(\frac{r_1}{r_2}\right)^n P_n(\cos{S}),
\end{equation} 
where $S$ is the angle between vectors ${\bf r_1}$ and ${\bf r_2}$. We choose to perform the development for all $P_n$ with $n \leq 3$. This well-known development, limited to order 3 in the semi-major axes ratio $\alpha=a_1/a_2$, is called {\it octupole theory} (see e.g. \citealt{For00}, \citealt{Lee03} and \citealt{Mig11}). For the sake of completeness, we present hereafter the technical details of the calculations.

Practically, the term $\cos{S}={\bf r}_1 \cdot {\bf r}_2/(r_1r_2)$ can be expressed in the following way - see \cite{Pra05} for a similar 2-D calculation:
\begin{eqnarray}
  \cos{S} &=& \left( R_1 \left(\begin{array}{c} \cos f_1\\ \sin f_1\\ 0 \end{array}\right) \right)^T \cdot R_2 \left(\begin{array}{c} \cos f_2\\ \sin f_2\\ 0 \end{array}\right)\\
  &\stackrel{R=R_1^T R_2}{=}& (\cos f_1\; \sin f_1\; 0) \, R \, \left(\begin{array}{c} \cos f_2\\ \sin f_2\\ 0 \end{array}\right)\\
  &=& \tilde\alpha \cos f_1 + \tilde\beta \sin f_1
\end{eqnarray}
where $R_1(i_1,\omega_1,\Omega_1)$ and $R_2(i_2,\omega_2,\Omega_2)$ are Eulerian rotations of the orbital reference frames of the mass $m_1$ and $m_2$ respectively, and $\tilde\alpha$ and $\tilde\beta$ have rather simple expressions: 
\begin{equation}\label{alpha}
\tilde\alpha=R_{1,1}\cos f_2 +R_{1,2}\sin f_2, \quad 
\tilde\beta =R_{2,1}\cos f_2 +R_{2,2}\sin f_2,
\end{equation}
$R_{i,j}$ being the element in the $i$th row and $j$th column of the matrix R. 
The averaging of the Hamiltonian (\ref{Pn}) over the short-period terms,
\begin{equation}
<H_{DP}>_{M_1,M_2} \, = \frac{1}{(2\pi)^2}\int_0^{2\pi} \int_0^{2\pi} H_{DP} \, dM_1 dM_2,
\end{equation}
is realized, in practice, over the eccentric anomaly $E_1$ of the perturbed body and over the true anomaly $f_2$ of the outer perturbing body. 
The intermediate results after the first averaging are the following:
\begin{eqnarray}
  <H_{DP0}>_{M_1} \hspace{-0.3cm} &=& \hspace{-0.3cm}- \frac{G\,m_1\,m_2}{r_2},\\
  <H_{DP1}>_{M_1} \hspace{-0.3cm}&=& \hspace{-0.3cm}\frac{3 G\,m_1\,m_2}{2a_2^2}\,\left(\frac{a_2}{r_2}\right)^2 
  a_1 e_1 \tilde\alpha,\\
  <H_{DP2}>_{M_1} \hspace{-0.3cm}&=& \hspace{-0.3cm} -\frac{G\,m_1\,m_2}{4\,a_2^3}\left(\frac{a_2}{r_2}\right)^3 
  a_1^2 \left[(12\tilde\alpha^2-3\tilde\beta^2-3)\, e_1^2 \right.  \nonumber\\ 
  & & \hspace{2.3cm}\left. + 3(\tilde\alpha^2+\tilde\beta^2)-2\right],\\
  <H_{DP3}>_{M_1} \hspace{-0.3cm}&=& \hspace{-0.3cm}\frac{5 G\,m_1\,m_2}{16\,a_2^4}\, \left(\frac{a_2}{r_2}\right)^4 a_1^3 e_1 \tilde\alpha \left[(20\tilde\alpha^2 - 15\tilde\beta^2 - 9) \, e_1^2 \right. \nonumber\\ 
  & & \hspace{2.7cm} \left. + 15 (\tilde\alpha^2 + \tilde\beta^2) - 12\right],
\end{eqnarray}
where $H_{DPi}$ means the term of the direct part (\ref{Pn}) associated to the $i$th Legendre polynomial. 
For the averaging over the true anomaly of the outer body, we first replace $\tilde\alpha$ and $\tilde\beta$ by their values (see Eq. (\ref{alpha})), and obtain the following first terms of the secondly averaged Hamiltonian:
\begin{eqnarray}
  <H_{DP0}>_{M_1,M_2} \hspace{-0.15cm}& = &\hspace{-0.15cm} -\frac{G\,m_1\,m_2}{a_2}, \\
  <H_{DP1}>_{M_1,M_2} \hspace{-0.15cm}& = &\hspace{-0.15cm} 0,
\end{eqnarray}
the terms $<H_{DP2}>_{M_1,M_2}$ and $<H_{DP3}>_{M_1,M_2}$ being too long to be exposed here. As we can see, the first two terms are constant in the secular problem and do not contribute to the averaged Hamiltonian~${\mathcal K}$. Let us note that an alternative development of the secular expansion using Hansen coefficients can be found in Laskar \& Bou\'e (2010). 

To simplify the formulation of the averaged Hamiltonian, the Jacobi's reduction, also known as the {\it elimination of the nodes} (\citealt{Jac42}), allows us to reduce the expansion to a two degree of freedom function only. The reduction is based on the existence of additional integrals of motion, namely the invariance of the total angular momentum, $\bf{C}$, both in norm and in direction. The constant direction of the vector $\bf{C}$ defines an invariant plane perpendicular to it. This plane is known as the {\it invariant Laplace plane}. The choice of this plane as reference plane implies the following relations:
\begin{eqnarray}
& & \Omega_{1}-\Omega_{2}=\pm180^\circ \label{cond1} \\
& & G_{1}\cos{i_{1}}  +  G_{2}\cos{i_{2}} = C  \label{cond2}\\
& & G_{1}\sin{i_{1}}  -  G_{2}\sin{i_{2}} = 0  \label{cond3}
\end{eqnarray}
with $C$ being the norm of the total angular momentum. Let us note that, in our study, the invariant Laplace plane coincides almost with the Jupiter-like planet's orbital plane, since its inclination relative to the invariant plane, $i_2$, is of the order of $m'_1/m'_2$ by relation (\ref{cond3}).

Another quantity, related to the total angular momentum, is frequently used. This is known as the {\it angular momentum deficit} (Laskar 1997):
\begin{equation}\label{AMD}
{\rm AMD} = \sum_{j=1}^2 L_{j} (1-\sqrt{1-e_{j}^2}\cos{i_{j}})=L_{1}+L_{2}-C.
\end{equation}

Finally, we present the octupole expansion of the Hamiltonian~(\ref{democratic}) averaged over the short-period terms and expressed in the invariant Laplace plane, using the succinct formulation introduced by \cite{For00}:
\begin{equation}\label{final}
\begin{array}{lll}
{\mathcal K}  = & - \;  \alpha^2 \; \frac{G m_1 m_2 L_2^3}{16 a_2 G_2^3} \;\,  & \left[\left(2+3\left(1-\left(\frac{G_1}{L_1}\right)^2\right)\right) \left(3\cos^2{i}-1\right)  \right.\\
& & \left. + 15 \left(1-\left(\frac{G_1}{L_1}\right)^2\right) \left(1-\cos^2{i}\right) \cos{2g_1} \right]\\ 
& +\; \alpha^3 \; \frac{15 G m_1 m_2 L_2^5}{64 a_2 G_2^5} \;  & \sqrt{1-\left(\frac{G_1}{L_1}\right)^2} \sqrt{1-\left(\frac{G_2}{L_2}\right)^2} \\
& &\left[ A (-\cos{g_1}\cos{g_2}-\cos{i}\sin{g_1}\sin{g_2}) \right.  \\
& & \left. +10 \left(\frac{G_1}{L_1}\right)^2 \cos{i} (1-\cos^2{i})  \sin{g_1} \sin{g_2}  \right]
\end{array}
\end{equation}
where 
$$\begin{array}{l}
\cos{i}=\frac{C^2-G_1^2-G_2^2}{2G_1G_2}\\
B=7-5\left(\frac{G_1}{L_1}\right)^2-7\left(1-\left(\frac{G_1}{L_1}\right)^2\right)\cos{2g_1}\\
A=7-3\left(\frac{G_1}{L_1}\right)^2-\frac{5}{2}(1-\cos^2{i})B,\\
\end{array}$$
$i=i_1+i_2$ being the mutual inclination. The equations of motion associated to Hamiltonian (\ref{final}) are 
\begin{equation}\label{eqfinal}
\dot{g}_i=\frac{\partial{\mathcal K}}{\partial G_i}, \quad  \dot{G}_i=-\frac{\partial{\mathcal K}}{\partial g_i}.
\end{equation}

One has to keep in mind that such an approach is limited to small values of the semi-major axes ratio, namely $\alpha<0.1$. To consider larger values of the ratio, a development to higher order is needed, as done by \cite{Koz62}.

For a Jupiter-like planet on a circular orbit ($G_2=L_2$), the formulation (\ref{final}) simplifies to the quadrupole approximation (second-order terms in $\alpha$). Then the secular Hamiltonian does not depend on $g_2$, and the norm of the associated momentum $G_2$ is an integral of motion, which means that the eccentricity of the outer body is constant in this formulation. As a result, the problem is integrable and this approximation is studied in many papers (e.g. \citealt{Har69}, \citealt{Lid76}, \citealt{Fer94} and \citealt{Far10}, or in the artificial satellite context e.g. \citealt{Lid62}, \citealt{Rus09} and \citealt{Del10}). 

In the present work, no assumption on the eccentricity of the Jupiter-like planet is considered. The variation of the eccentricity of the perturber is introduced through the octupole terms (third-order terms in $\alpha$). As these terms depend on the variable $g_2$, the problem can not be reduced to one degree of freedom, and it induces qualitative changes on its dynamics, as shown in the next section.

\section{Geometric representation of the dynamics}

In this section, we study the dynamics of the two degree of freedom Hamiltonian (\ref{final}) by means of 2-D geometric representations, called {\it representative planes} (see \citealt{Mic06}, \citealt{Lib07}). The idea is to choose a 2-D plane of initial conditions which is suitable for the analysis of the stationary solutions of the secular two degree of freedom problem. This plane should be representative in the sense that we aim to find a plane such that it contains the initial conditions of orbits representative of each class of orbits. 

Such a plane can be obtained by fixing $g_1$ and $g_2$ to values that verify the conditions 
\begin{equation}\label{cond}
\dot{G}_1=\frac{\partial{\mathcal K}}{\partial g_1}=0 \quad {\rm and} \quad  \dot{G}_2=-\frac{\partial{\mathcal K}}{\partial g_2}=0,
\end{equation}
i.e. according to the symmetries of the secular 3-D problem, $(2 g_1,\Delta \omega=g_1-g_2)=(0,0), (0,\pi), (\pi,0)$ and $(\pi,\pi)$. Indeed, the secular Hamiltonian function can be developed in Fourier series of cosinus terms, whose generic argument is 
\begin{equation}
\phi=j_1g_1+j_2g_2+k\Delta\Omega,
\end{equation}
where $j_1$ and $j_2$ are of the same parity ($j_1, j_2, k$ are integers), and $\Delta\Omega=\pi$ after node reduction. As a result, conditions (\ref{cond}) are verified when $\sin{\phi}=0$, i.e. $(2 g_1,\Delta \omega)=(0,0), (0,\pi), (\pi,0)$ and $(\pi,\pi)$. These four pairs of angles define four distinct quarters of the representative plane. 

In the following, we choose the geometric representation introduced by \citet{Mig11}, and defined as $x=e_1 \cos{2g_1}$ and $y=e_2\cos{\Delta \omega}$ with $\sin{2g_1}=\sin{\Delta \omega}=0$. On this representative plane, the level curves of constant Hamiltonian are plotted for given values of AMD, $\alpha=a_1/a_2$ and $\mu=m_1/m_2$. The boundary of permitted motion is defined as $i=0,180^\circ$ hereafter. Let us recall that the eccentricities and mutual inclination are related through the integral of AMD. 

We insist on the fact that the representative plane is not a phase space or a surface of section. However, all orbits have to cross the representative plane (i.e. pass through the conditions  $\sin{2g_1}=\sin{\Delta \omega}=0$ whatever the behavior of the angles $2g_1$ and $\Delta \omega$), and the points of intersection have to follow a constant energy curve. As the extremal values of the eccentricities are reached when $\sin{2g_1}=\sin{\Delta \omega}=0$ (\citealt{Mic06}, \citealt{Lib08}, \citealt{Lib09}), a quasi-periodic solution intersects the representative plane at four points on the same energy level. A stationary solution, fulfilling the two additional conditions $\dot{g}_i={\partial{\mathcal K}}/{\partial G_i}=0$ ($i=1,2$), appears as a fixed point on the plane, while a periodic solution for which an angle is fixed has only two points of intersection. Orbits of chaotic motion intersect it at an arbitrary number of points. Depending on the location of these intersection points on the four quadrants of the plane, the behavior of the angles can also be deduced, as well as an estimation of the ranges of eccentricity variations, as will be shown in the following examples. 

The mass ratio $\mu$ is fixed to $10^{-4}$ in the following. Concerning the semi-major axis ratio, the use of the octupole terms limits the width of semi-major axes ratios that can be considered; to ensure the validity of our approach, we choose $\alpha=0.05$. Indeed, \cite{Mig11} have shown that, for hierarchical systems, the octupole formulation is very precise and higher-order contributions do not distort the structure of the Hamiltonian curves of the representative plane.  

Two initial configurations of the three-planet system are examined in this work. In Section \ref{circ}, the outer giant perturber is considered on a nearly circular orbit, while the influence of a highly elliptic orbit of the perturber is analyzed in Section \ref{ell}.

\begin{figure*}
\includegraphics[height=6cm]{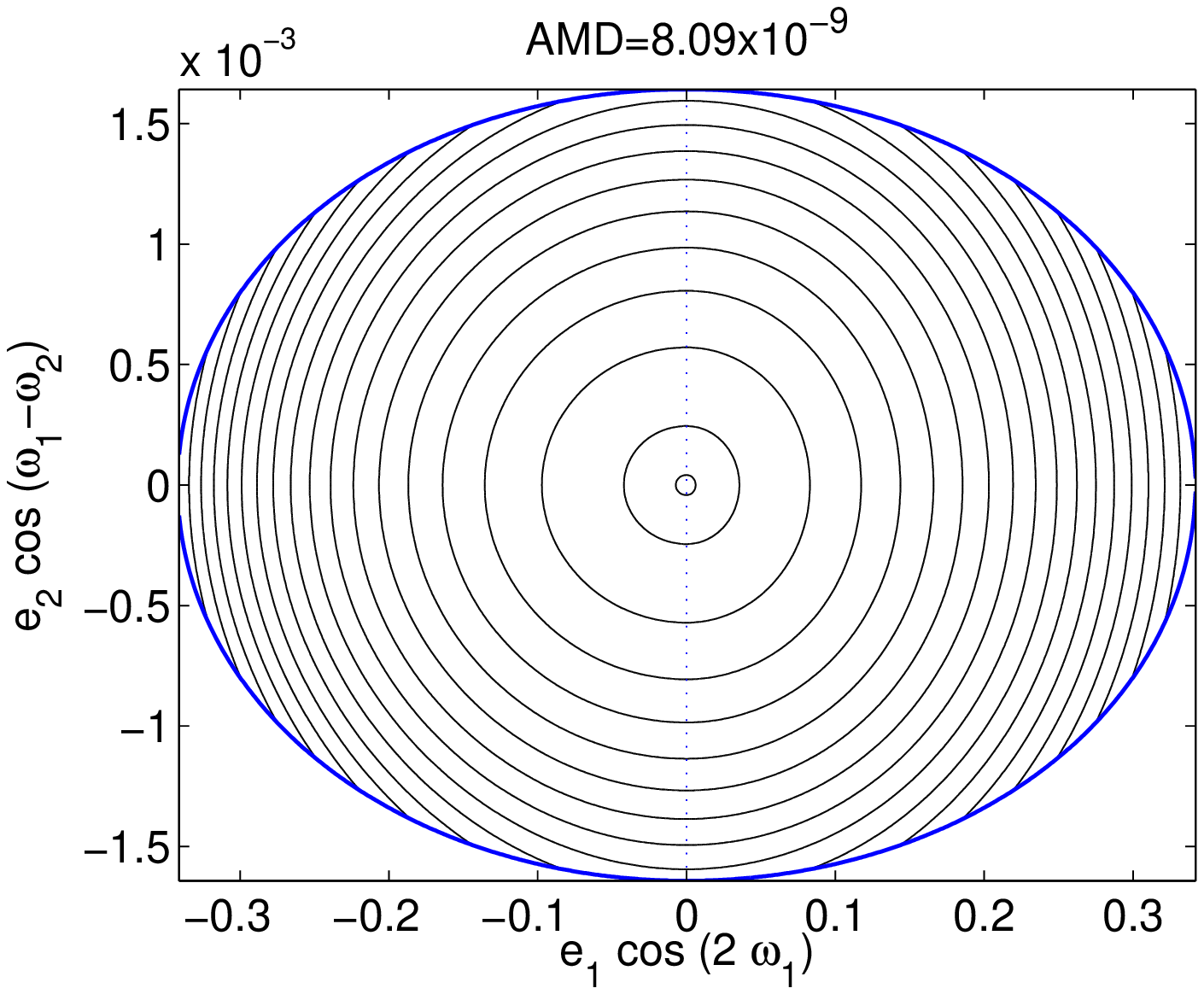}
\includegraphics[height=6cm]{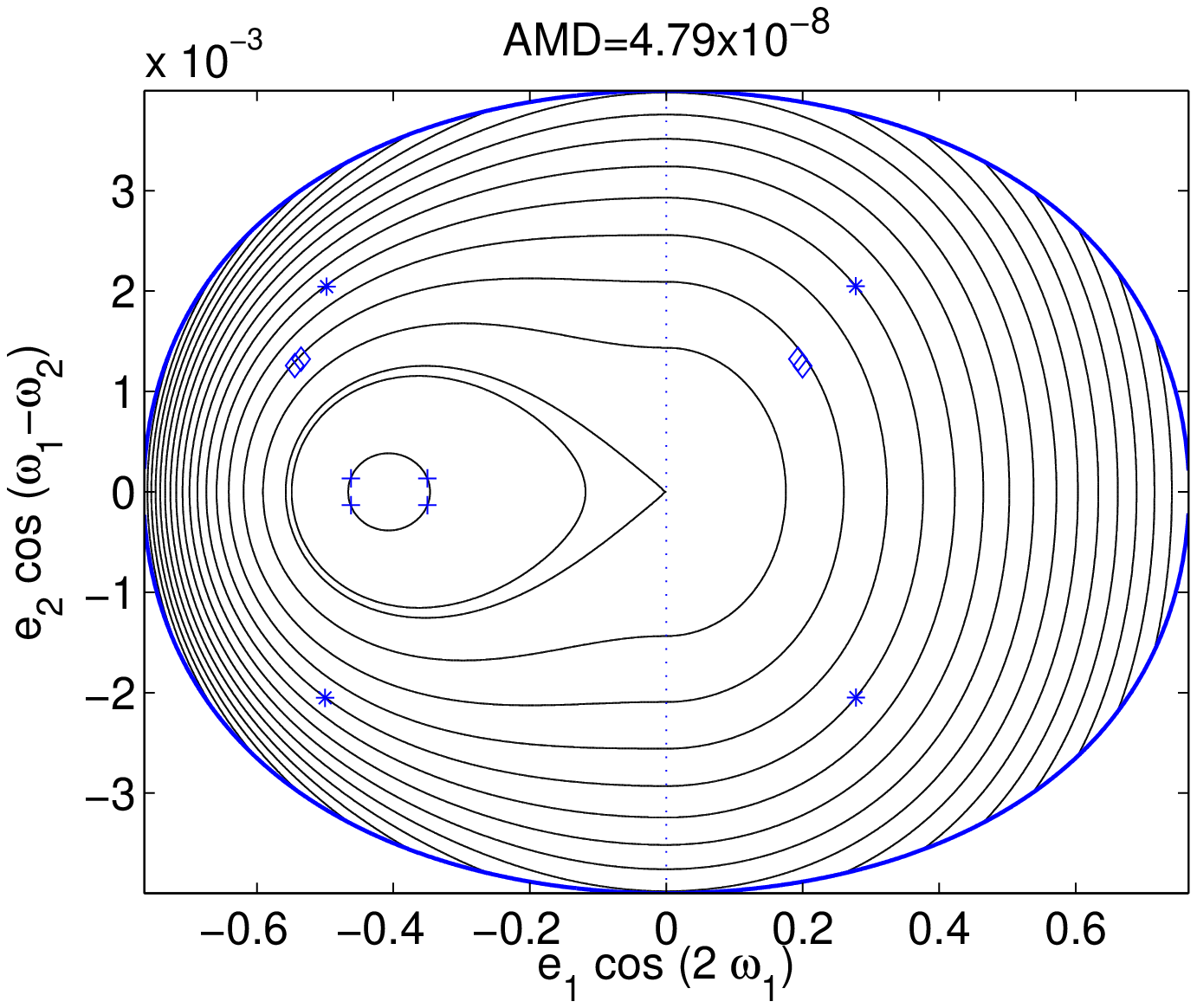}
\includegraphics[height=6cm]{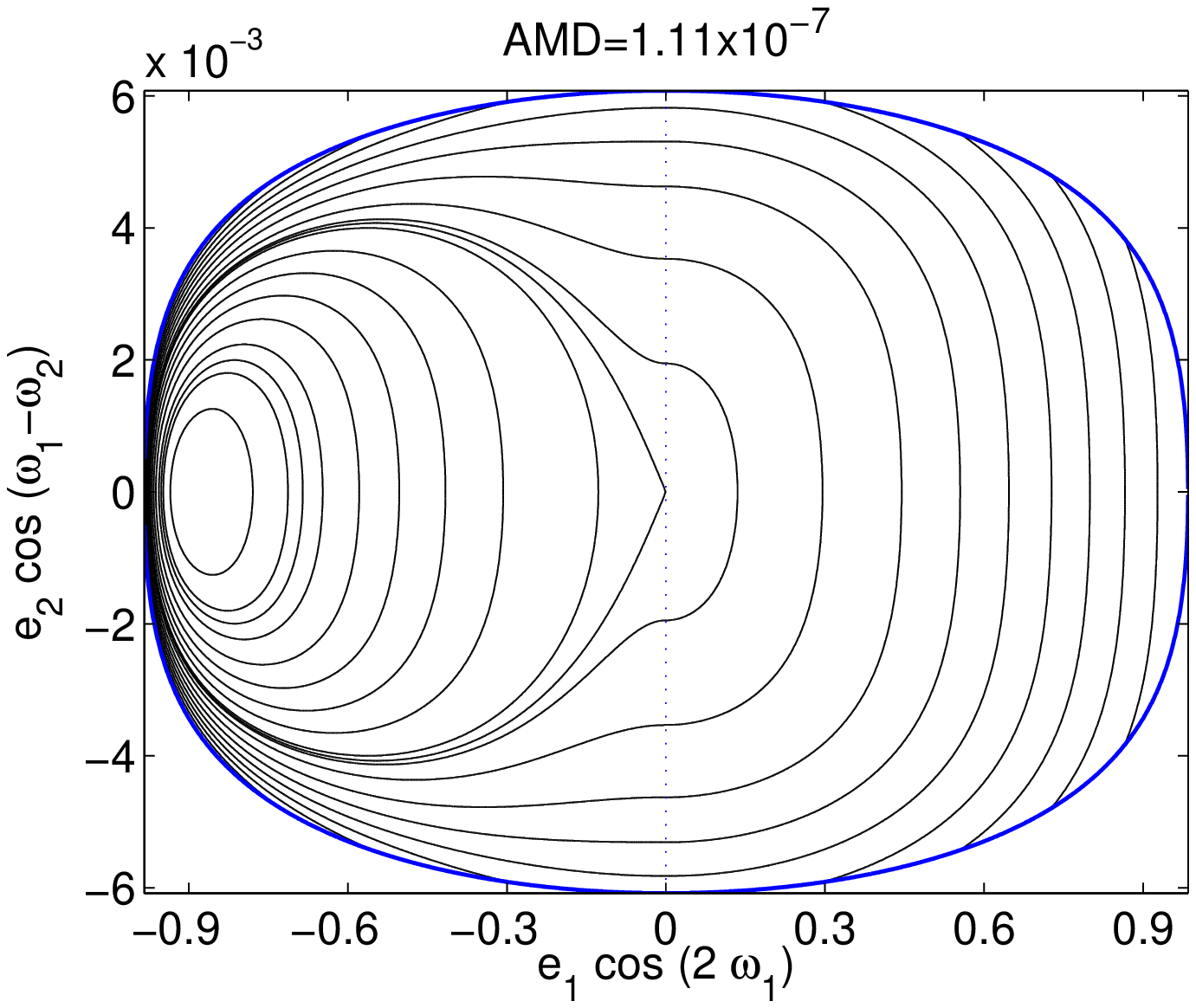}
\includegraphics[height=6cm]{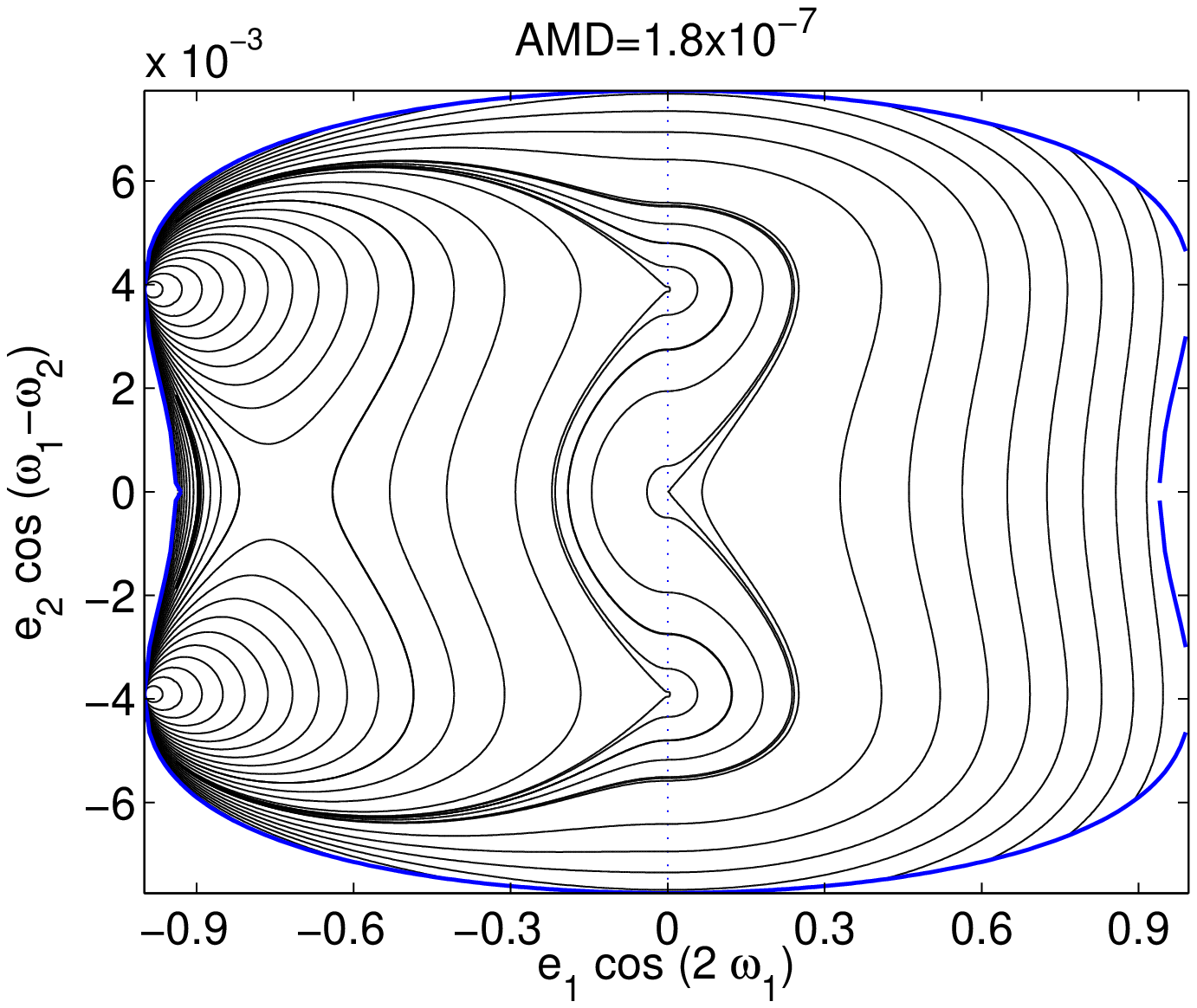}
\includegraphics[height=6cm]{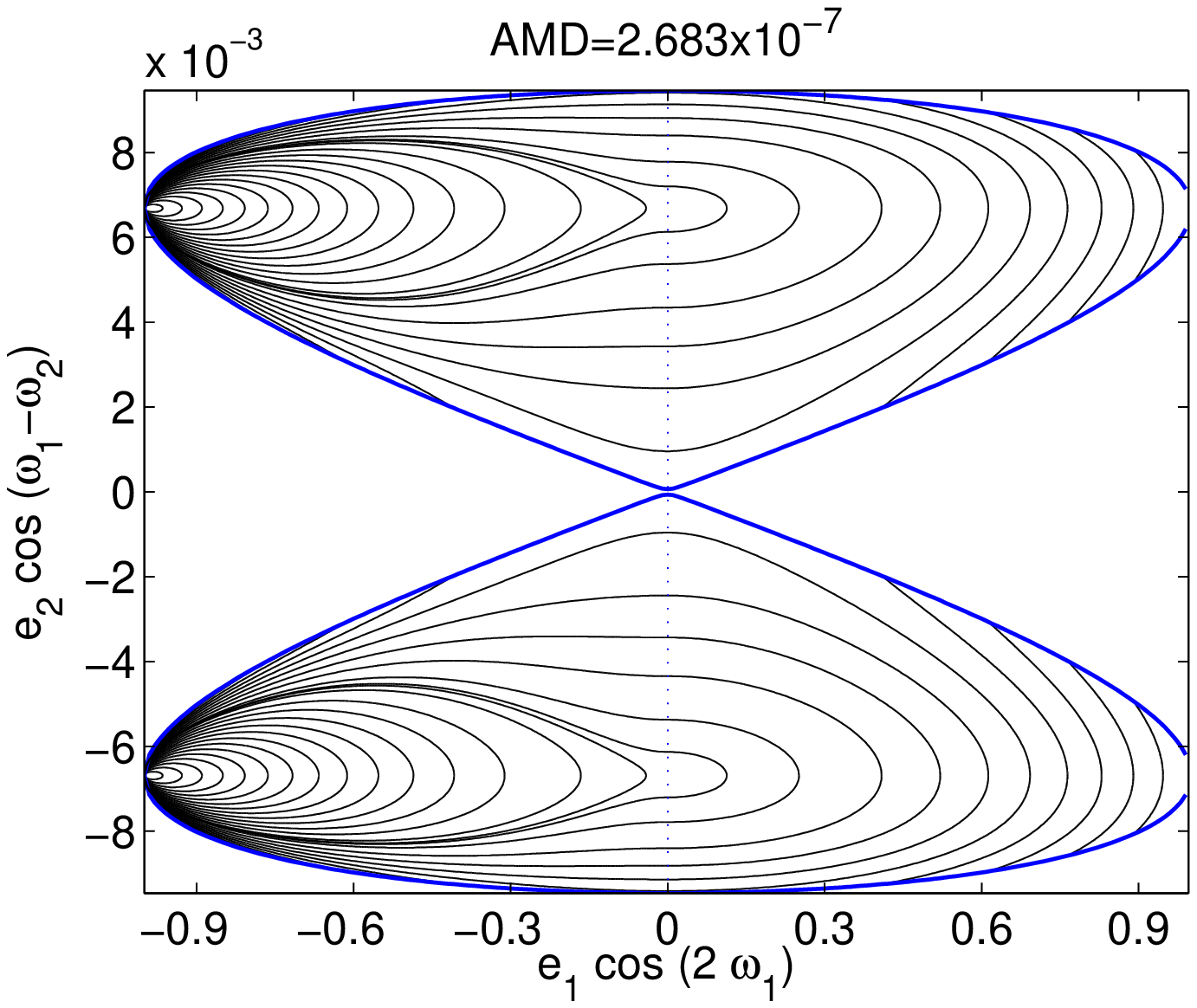}
\includegraphics[height=6cm]{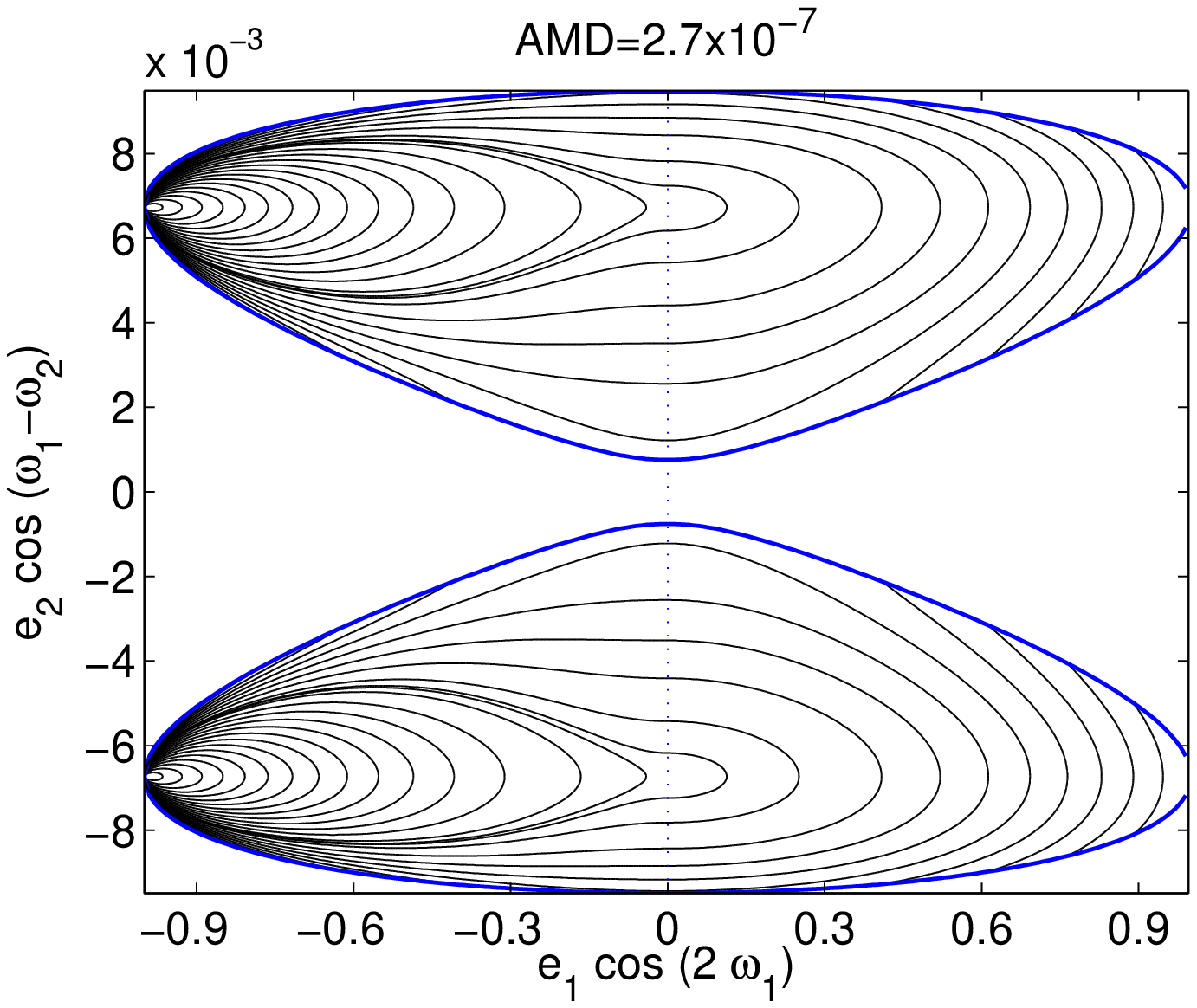}
\caption{Level curves of constant Hamiltonian (\ref{final}) in the representative plane $(e_1 \cos{2 g_1}, e_2 \cos{\Delta \omega})$ for different values of AMD such that the mutual inclination at the origin is $20^\circ$ (top left), $50^\circ$ (top right), $80^\circ$ (middle left), $110^\circ$ (middle right), $180^\circ$ (bottom left). In the bottom right panel, the origin $e_1=e_2=0$ does not belong to the region of permitted motion. Other parameters are $\alpha=0.05$ and $m_1/m_2=10^{-4}$.}
\label{espcirc}
\end{figure*}

\begin{figure*}
\centering
\includegraphics[width=5.8cm]{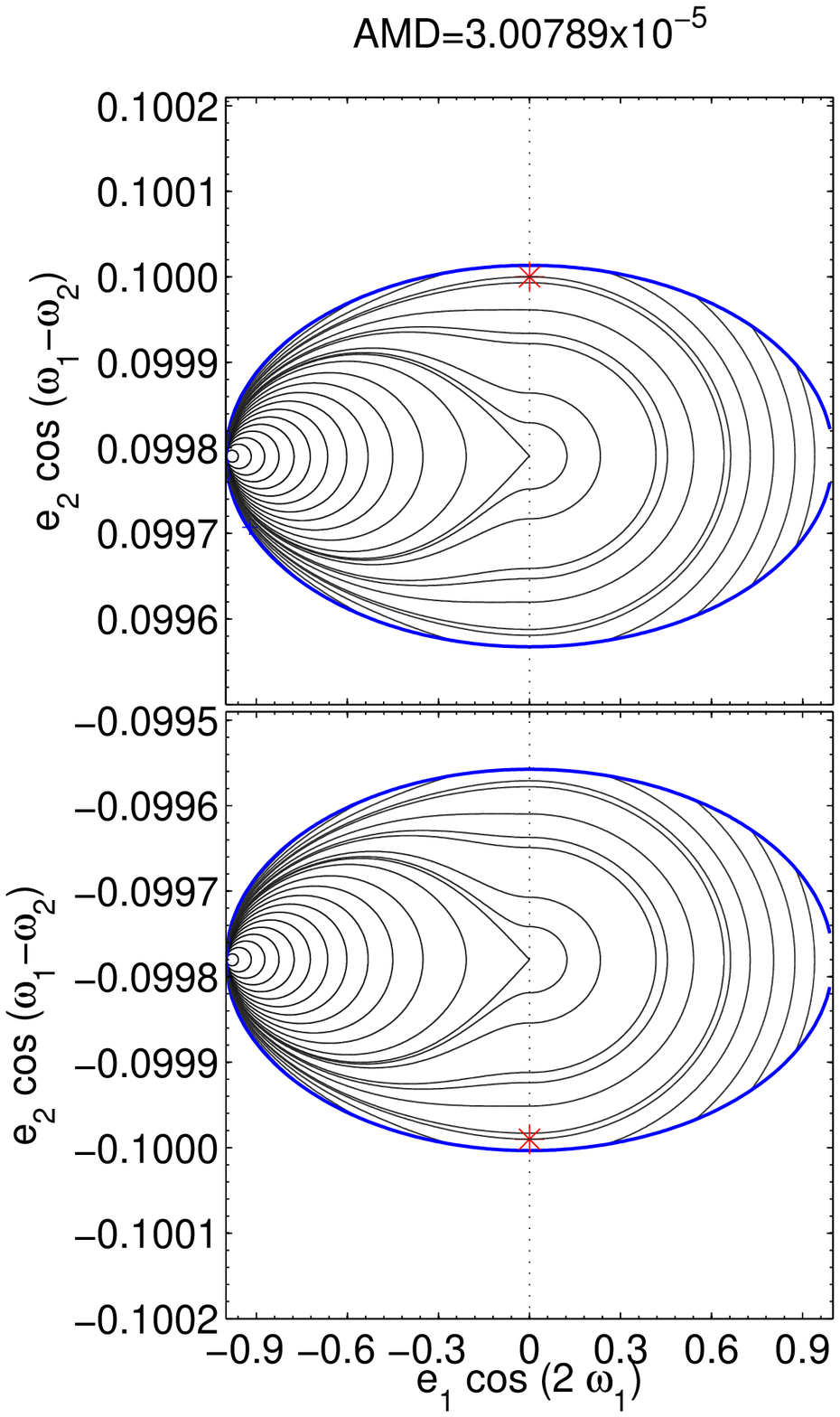}
\includegraphics[width=5.8cm]{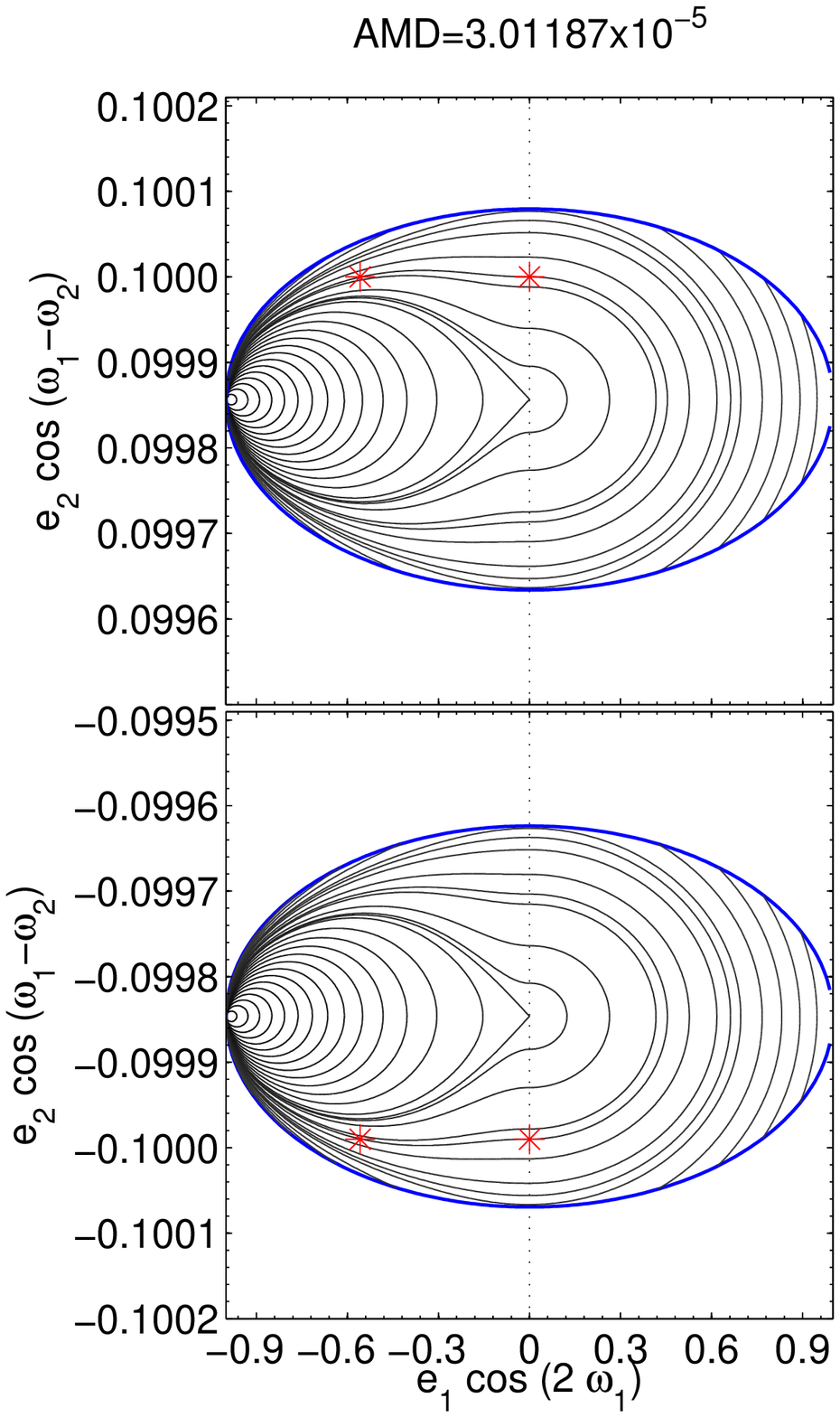}
\includegraphics[width=5.8cm]{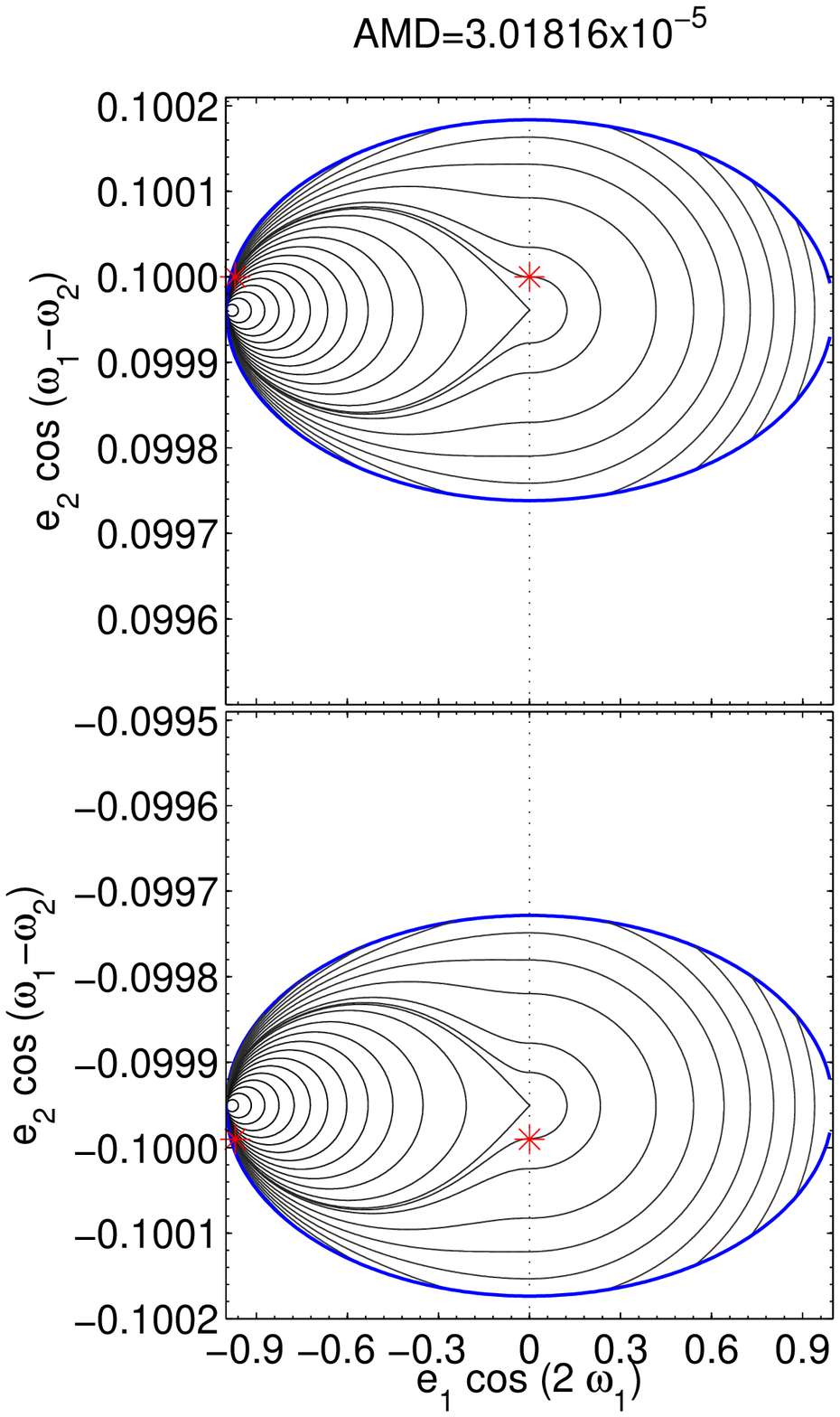}
\caption{Level curves of constant Hamiltonian (\ref{final}) in the representative plane $(e_1 \cos{2 g_1}, e_2 \cos{\Delta \omega})$ for different values of AMD such that the mutual inclination of the orbits with initial eccentricities $e_1=0$ and $e_2=0.1$ is $20^\circ$ (left), $50^\circ$ (middle) and $80^\circ$ (right). Other parameters are $\alpha=0.05$ and $m_1/m_2=10^{-4}$.}
\label{esp01}
\end{figure*}

\subsection{Nearly circular orbit of the perturber}\label{circ}

When the perturbing body is on a circular orbit, the octupole formulation coincides with the quadrupole approximation. Since the quadrupole approach ${\mathcal K}_{quad}(e_1,\omega_1)$ is integrable, its dynamics can be represented on the phase space $(e_1 \cos{\omega_1}, e_1 \sin{\omega_1})$ (see e.g. \citealt{Tho96} for more detail). However, for nearly circular orbit of the perturber, the expansion (\ref{final}) is four-dimensional and a first picture of the dynamics consists in the plot of the level curves of constant Hamiltonian in the aforementioned representative plane. This representation is given in Figure \ref{espcirc} for several values of AMD: $8.09\times10^{-9}$, $4.79\times10^{-8}$, $1.11\times10^{-7}$, $1.80\times10^{-7}$, $2.683\times10^{-7}$ and $2.7\times10^{-7}$. For all these values expect the last one, the {\it maximal mutual inclination} between the two orbital planes, $i_{max}$, is reached at the origin ($e_1=e_2=0$), while the border enclosing the possible dynamics of the problem represents the coplanar case ($i=0$). The five first AMD values considered here correspond to $i_{max}$ of $20^\circ$, $50^\circ$, $80^\circ$, $110^\circ$ and $180^\circ$ respectively. In the bottom right panel of Figure \ref{espcirc}, the region of permitted motion separates into two parts which are bordered by the curves $i=0^\circ$ (higher absolute values of $e_2$) and $i=180^\circ$ (smaller absolute values of $e_2$). 

As explained hereabove, the structure of the geometric representation reveals the equilibria of the problem. For small mutual inclination ($i_{max} = 20^\circ$, see Figure \ref{espcirc} top left), circular orbit of the inner body corresponds to a stable equilibrium and no variation in eccentricity is possible. For larger inclinations ($i_{max} = 50^\circ$ and $80^\circ$, top right and middle left panels respectively), the point $e_1=0$ becomes an unstable equilibrium, and a separatrix divides the left panel of the representative plane (where $2g_1=\pi$): the closed region is characterized by the libration of $g_1$ around $90^\circ$ or $270^\circ$ and the region outside the separatrix by the circulation of this angle. The two stable equilibria (at $g_1=90^\circ$ and $g_1=270^\circ$) created by bifurcation of the equilibrium at circular orbit are referred to as {\it Kozai equilibria}, by analogy to the restricted problem (Kozai 1962; Lidov 1962). This change of stability of the central equilibrium induces large variation in eccentricity for an inner body initially on a nearly circular orbit, since its real motion (short periods included) will stay close to the separatrix of the reduced problem. The maximal mutual inclination corresponding to the change of stability of the central equilibrium, called {\it critical mutual inclination}, has been calculated by \cite{Koz62}: it drops from the well-known value $39.23^\circ$ to $32^ \circ$, as the semi-major axes ratio increases from $0$ to $0.5$. For the parameters of Figure \ref{espcirc}, the critical mutual inclination is $39.1^\circ$. 

Additional bifurcations of these equilibria occur for higher values of mutual inclination (see middle right panel of Figure~\ref{espcirc}). For more detail, we refer to the complete study of these equilibria and their stability realized by \citet{Mig09} for the three-body problem. For increasing values of AMD, the equilibrium at the origin vanishes and the region of permitted motion is divided into two islands. The dynamics is then governed by two families of equilibria: the equilibria related to the bifurcation of the Kozai equilibria and located at the border of permitted motion (called solutions IVa by \cite{Mig09}), and those related to the bifuracion of the central equilibrium and appearing close to the $e_1=0$ axis (called solution IIIa by \cite{Mig09}). Let us note that these last ones are unstable.


In the Laplace plane reference frame, \citet{Lib08} have shown that, when the orbit is outside the Kozai-resonant area, the global extrema of the eccentricities are reached when $\sin{\Delta \omega}=0$ (see also \citealt{Mic06}), while their local extrema are reached when $\sin{2g_1}=0$. Example of such a behavior is illustrated in Fig \ref{espcirc} (top right panel) where the four points of intersection of a given orbit are symbolized by '*' signs. As they are located in the four quadrants, the dynamics of this orbit is characterized by the circulation of both angles $2g_1$ and $\Delta \omega$. For a Kozai-resonant system considered in the Laplace plane reference frame, \citet{Lib09} have shown that the eccentricities of both planets are not coupled, the eccentricity of the inner planet being extremal when $\sin{2g_1}=0$, and the one of the outer planet when $\sin{\Delta \omega}=0$. The '+' signs in Fig \ref{espcirc} (top right panel) show a Kozai-resonant orbit: all the intersection points are located on the left part of the representation, indicating the libration of $g_1$ resonant angle.

For the reasons given above, a particular interest of such a geometric view of the dynamics is to give an estimation of the variation in eccentricity of each body. Let us note that, due to our choice of mass ratio $\mu$, the eccentricity of the outer massive body is only weakly affected by its small inner companion. Indeed, the long-term variation in eccentricity is described by the Hamiltonian equation (\ref{eqfinal}):
\begin{equation}
 \dot e_i=\frac{\sqrt{1-e_i^2}}{L_i e_i} \frac{\partial{\mathcal K}}{\partial g_i}.
\end{equation}
Thus the variation of the outer eccentricity is of the order of $m_1$, which is very small in this work, while the variation of the small body's eccentricity is quite important, as it is of the order of $m_2$. As a result, the eccentricity of the perturbing body is nearly constant and it explains that all the intersection points of an orbit seem to have the same ordinate (in absolute value) in Fig. \ref{espcirc} (top right panel). On the contrary, the variation in eccentricity of the inner body can be very significant. For instance, the eccentricity of the orbit denoted by '*' in Fig \ref{espcirc} (top right panel) varies roughly from 0.28 (positive abscissa) to 0.5 (negative abscissa).

The representations of Figure \ref{espcirc} give information on the dynamics of a system with nearly circular orbit of the perturber ($e_2$ smaller than $0.01$). In the next section, dynamics with higher initial values of $e_2$ will be considered.

\begin{figure*}
\centering
\rotatebox{270}{\includegraphics[width=12cm]{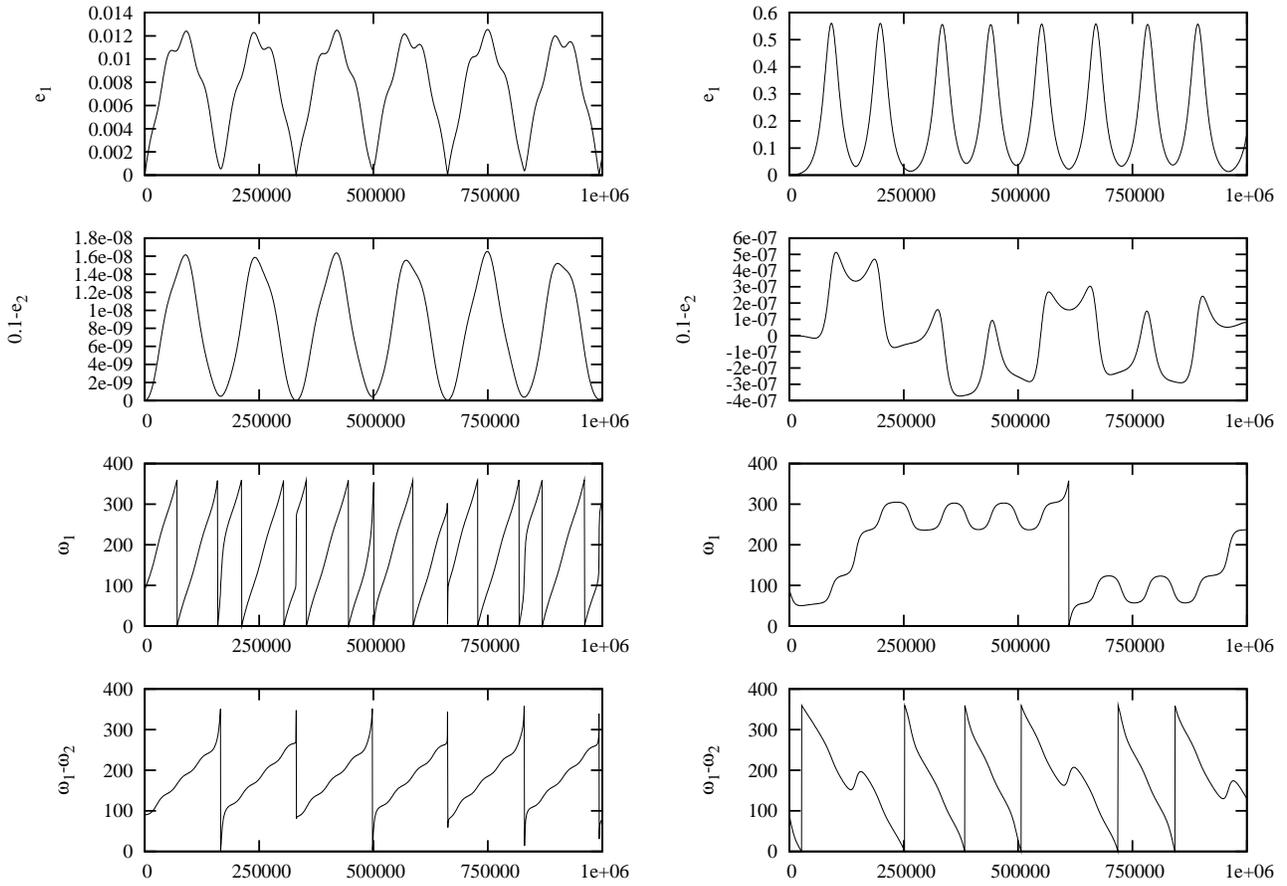}}
\caption{Long-term evolution of a system consisting of a small body initially on nearly circular orbit ($e_1=10^{-6}$) and an outer body whose initial eccentricity is $e_1=0.1$. Arguments of pericenter are fixed to $g_i=0^\circ$. The initial mutual inclination between both orbital planes is $i=20^\circ$ (left panel) and $i=50^\circ$ (right panel). The change of dynamics is obvious.}
\label{figlgterme}
\end{figure*}

\begin{figure*}
\centering
\includegraphics[width=14.5cm]{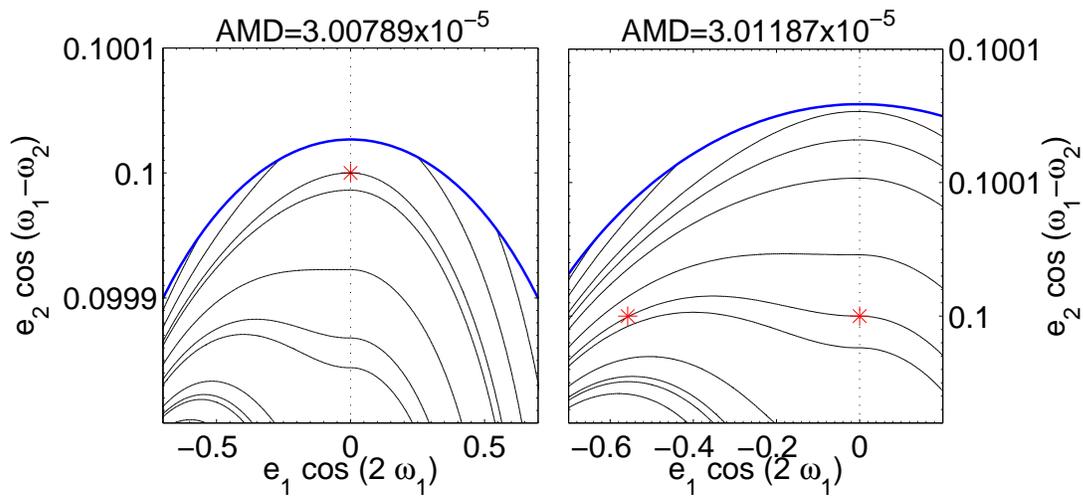}
\caption{Detailed views of Figure \ref{esp01}.}
\label{figzoom}
\end{figure*}

\begin{figure*}
\centering
\includegraphics[width=5.8cm]{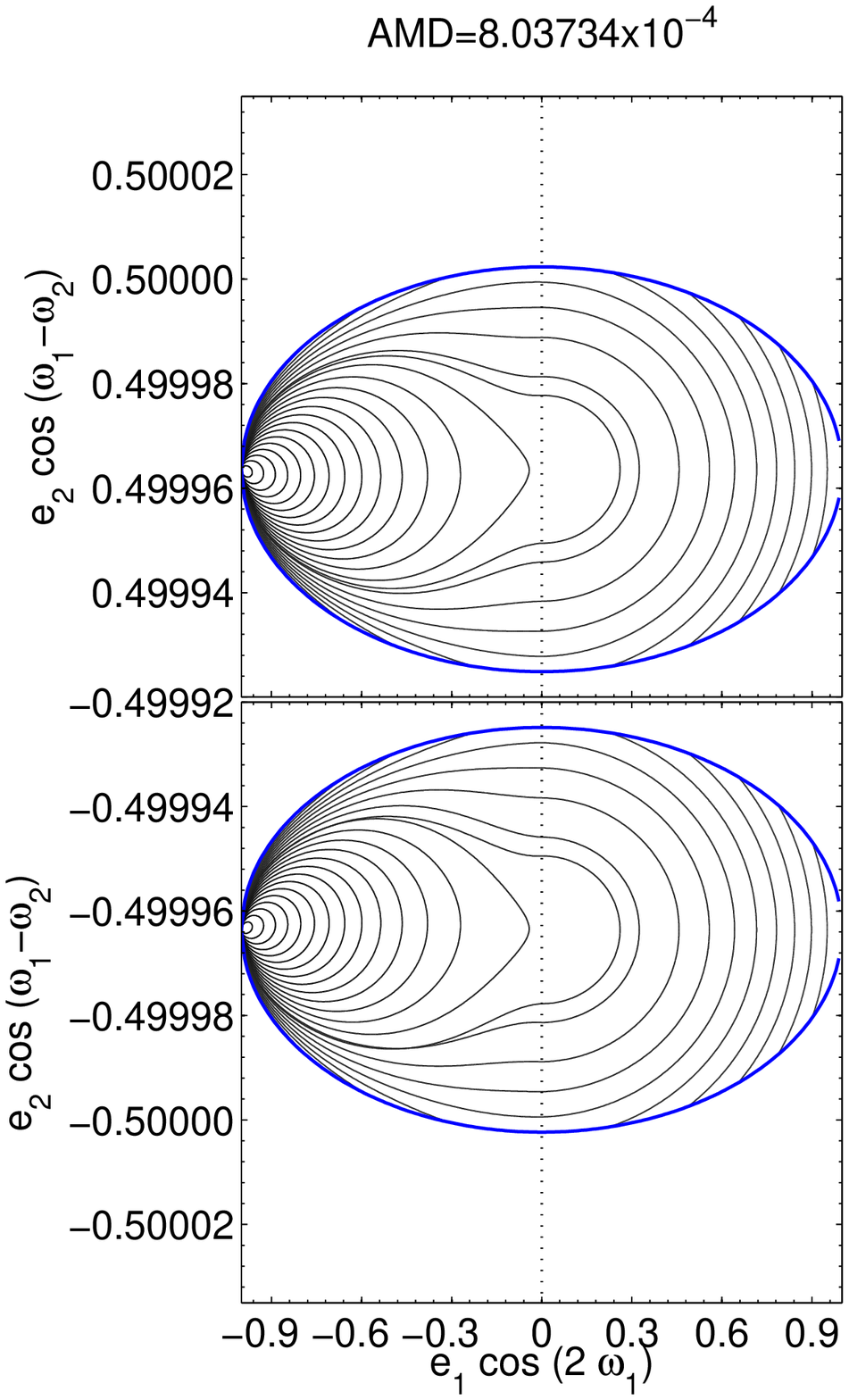}
\includegraphics[width=5.8cm]{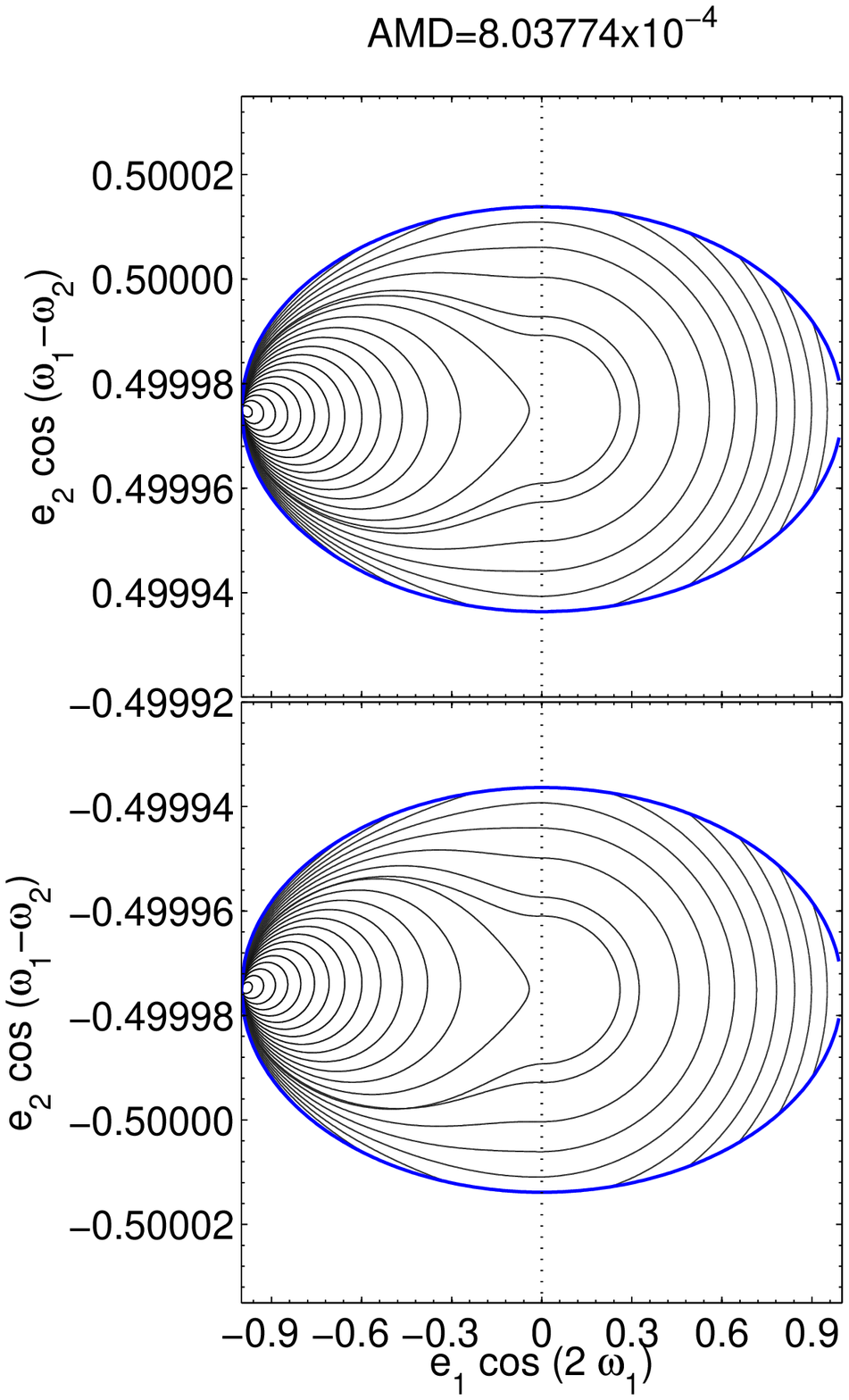}
\includegraphics[width=5.8cm]{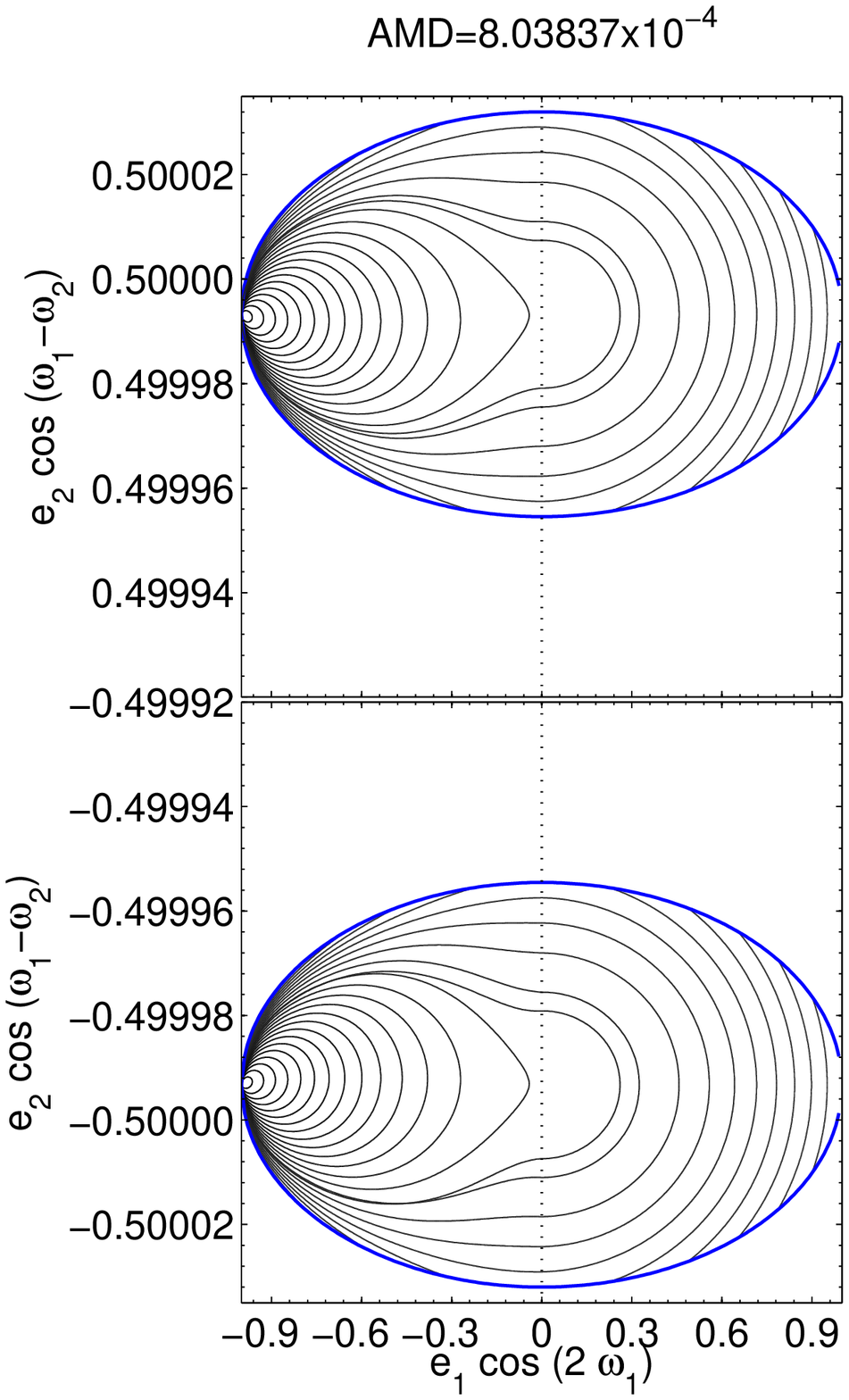}
\caption{Same representation as Figure \ref{esp01} for AMD values such that the mutual inclination of the orbits with initial eccentricities $e_1=0$ and $e_2=0.5$ is $20^\circ$ (left), $50^\circ$ (middle) and $80^\circ$ (right).}
\label{esp05}
\end{figure*}

\subsection{Elliptic orbit of the perturber}\label{ell}

In this section we assume an elliptic orbit of the perturber and study the dynamical evolution of the system by means of the representative plane. While the dynamics of Figure \ref{espcirc} focus on nearly circular orbit of $m_2$, we keep increasing the AMD values to reach higher eccentricities of this body. First let us consider $e_2$ close to $0.1$. Three different values of AMD are displayed in Figure \ref{esp01}: they are chosen such that the mutual inclination of the orbits with initial eccentricities $e_1=0$ and $e_2=0.1$ is $20^\circ$ (left), $50^\circ$ (middle) and $80^\circ$ (right). Our first observation is the similarity to the dynamics of the bottom right panel of Figure \ref{espcirc}. So the 3-D elliptic three-body problem is affected by two kinds of equilibria only: the ``Kozai-bifurcated'' equilibria at very high value of $e_1$ and the equilibria at circular orbit of the inner body.

Although the dynamics is very similar for the three values of AMD displayed in Figure \ref{esp01}, the shifting on the y-axis is obvious and explains the different dynamics observed for a given system considered at various mutual inclinations. In order to explain analytically the results of \cite{Fun11} (behavior of an Earth-like body initially on an inner circular orbit in a gas-giant system), let us consider the evolution of a two-planet system whose initial eccentricities are $e_1=10^{-6}$ and $e_2=0.1$. The intersection points of the evolution of the system with the representative plane are denoted by '*' signs in Figure \ref{esp01}.

For a small mutual inclination ($i=20^\circ$), $g_1$ and $\Delta \omega$ circulate and both variations in $e_1$ and $e_2$ are so limited that the four expected intersection points seem gathered on two points only (see left panel of Figure \ref{esp01}). For a mutual inclination of the orbits of $50^\circ$ (middle panel), the system is destabilized by the unstable equilibria: $g_1$ vacillates between libration and circulation and $e_1$ reaches values as high as $0.55$. The same instability is present for the orbit of the third panel of Figure \ref{esp01} ($i=80^\circ$), where $e_1$ reaches a value close to 1. If we had extended the integration time, the intersection points would not be regular anymore, showing the chaotic evolution of highly inclined systems due to the closeness to the ``separatrix''. These two different long-term evolutions ($i=20^\circ$ and $i=50^\circ$) are illustrated in Figure~\ref{figlgterme}, by means of a numerical integration of the octupole Hamiltonian equations (\ref{eqfinal}).


This change of dynamical behavior can be easily deduced from the shape of the Hamiltonian curves on the representative plane, as it can be seen on Figure \ref{figzoom}. For small mutual inclination (i.e. close to the borders of higher absolute values of $e_2$), the systems whose inner orbit is circular correspond to extrema of the Hamiltonian curve (left panel of Figure \ref{figzoom}). On the other hand, for mutual inclinations higher than a value close to $40^\circ$, there exists another intersection point, belonging to the same curve of constant Hamiltonian, and of same eccentricity $e_2$ (right panel of Figure \ref{figzoom}); the abscissa of these intersection points represent the secular variation of the eccentricity of $m_1$. For a higher eccentricity of the outer body, the dynamics is similar, as shown in Figure \ref{esp05} ($e_2=0.5$).

So we conclude that an inner small body on a quasi-circular orbit attracted by a giant companion on an elliptic orbit behaves secularly in a similar way as in the circular three-body problem: small periodic variation of its eccentricity when the mutual inclination of the orbital planes is small, on the contrary to the large chaotic variation observed for mutual inclinations higher than a critical value of $\sim40^\circ$. These analytical results are consistent with the numerical study of \cite{Fun11}.

However, it is interesting to note that, even if the representative planes of Figures \ref{espcirc}, \ref{esp01} and \ref{esp05} precisely depict the dynamics around the central and Kozai families of equilibria, some additional dynamical features can be ``hidden''. For instance, the '$\Diamond$' symbols in Figure \ref{espcirc} (top right panel) identify an orbit characterized by a libration of the angle $\Delta \omega$ around $180^\circ$ (and simultaneous circulation of $g_1$). This kind of behavior is classified as mode 2 by \citet{Mic06} and its existence can not be deduced from the analysis of our representative plane. In the same way, no particular dynamics associated to $\sim35^\circ$ of mutual inclination, as the one reported by \cite{Fun11} and described in the next section, is visible on the representative planes of Figure \ref{esp01}.

\begin{figure*}
\centering{
\includegraphics[width=18cm]{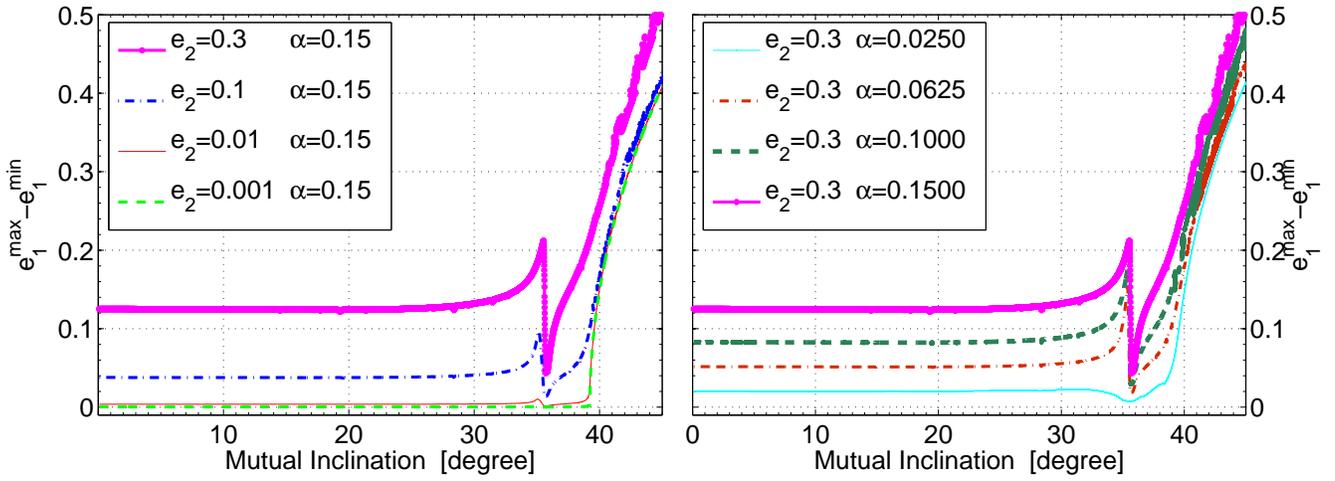}
}
\caption{Maximal eccentricity variation of the inner small planet on quasi-circular orbit ($e_1=10^{-6}$) reached during its secular evolution, for mutual inclination of the orbital planes up to $50^\circ$. The integration time is fixed to $10^7$ years.}
\label{35}
\end{figure*}

\begin{figure*}
\centering{
{\includegraphics[height=12cm]{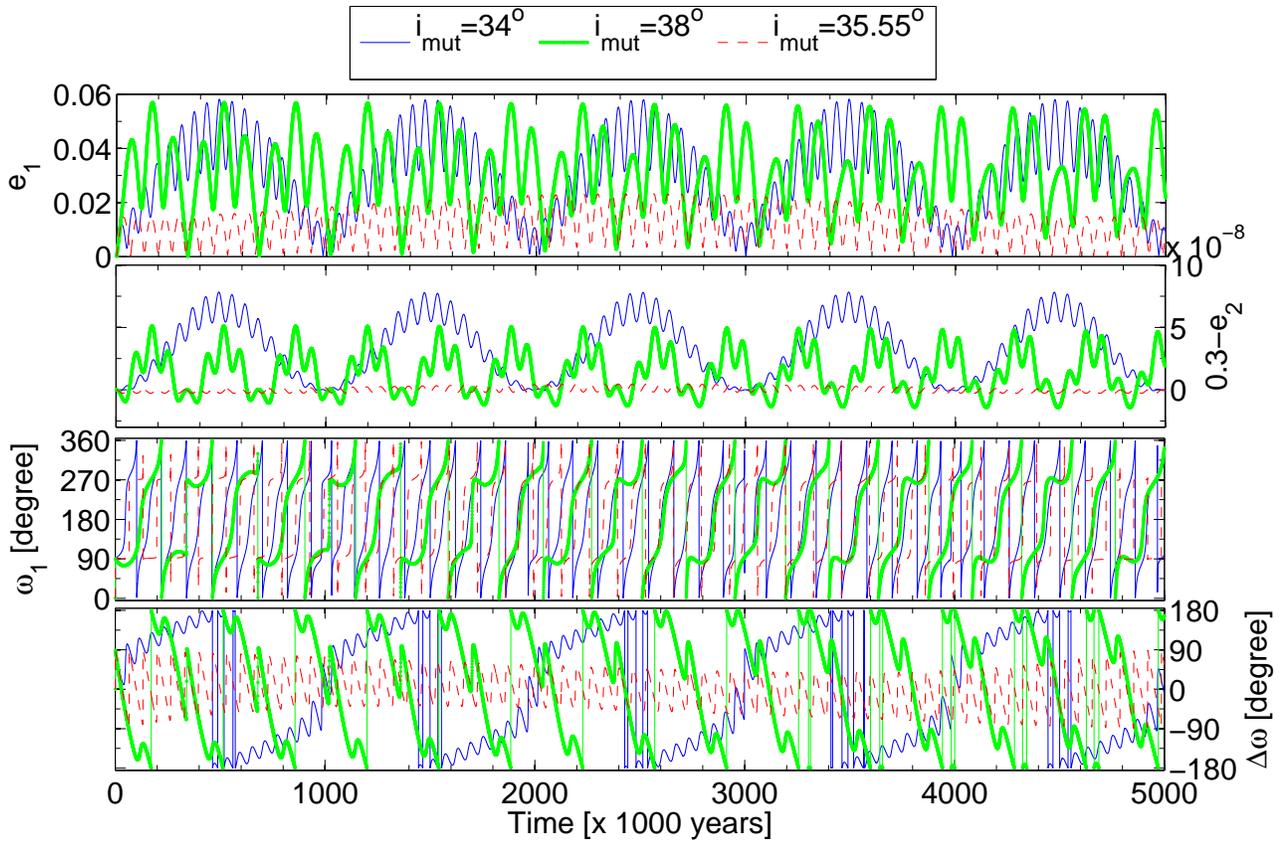}}
}
\caption{Long-term evolution of systems with mutual inclination of $34^\circ$, $35.55^\circ$ and $38^\circ$. Other initial orbital elements are: $a_1=0.05$, $a_2=1$, $e_1=10^{-6}$, $e_2=0.3$, $g_1=0^\circ$ and $g_2=0^\circ$. The masses are $m_1=10^{-4}M_{Jup}$ and $m_2=1M_{Jup}$.}
\label{lgterme}
\end{figure*}

\section{Interesting dynamics around $35^\circ$ of mutual inclination}

\citet{Fun11} have realized a numerical study of the long-term stability of inclined fictitious Earth-mass planets moving under the attraction of an eccentric giant planet. The small body is initially on a quasi-circular orbit. Although the massless companion exhibits periodically and limited variation in eccentricity for a mutual inclination smaller than $\sim 40^\circ$ (observation in agreement with our analytic study of the previous section), their simulations have identified a dynamical region around $35^\circ$ of mutual inclination consisting of long-time stable and particularly low eccentric orbits. 

This feature is well described by our octupole approximation, as it is shown in Figure \ref{35} where numerical integrations of the Hamiltonian equations (\ref{eqfinal}) are used to deduce the maximal eccentricity variation of the inner body on initially quasi-circular orbit ($e_1=10^{-6}$), and for mutual inclination of the orbital planes up to $50^\circ$. Left panel shows that, for small values of $i$, the variation of $e_1$ is negligeable when the orbit of the perturber is quasi-circular (due to the presence of the central equilibrium of Figure \ref{espcirc}), while, for eccentric orbit of the perturber, the higher the value of $e_2$ the wider the secular variation of $e_1$. This variation is even wider for high semi-major axes ratios, as shown in the right panel of Figure \ref{35}. For mutual inclinations higher than $\sim 38^\circ$, the instability related to the Kozai bifurcations and described in the previous section produces the important increase of the secular variations of $e_1$ (see the right sides of the graphs). The main new feature of Figure \ref{35} is the sudden decrease of the maximal $e_1$ variation around a value close to $35^\circ$ for all semi-major axes ratios and eccentricities of the gas giant, as observed in the numerical study of \cite{Fun11}. Let us note that this region around $35^\circ$ is not present for a circular orbit of the perturber and is much in evidence for high semi-major axes ratios, so that the change of dynamics is due to the third-order terms of the octupole expansion. The aforementioned behavior is also illustrated, in Figure \ref{lgterme}, by means of long-term evolutions of systems with mutual inclination of $34^\circ$, $35.55^\circ$ and $38^\circ$ (numerical integration of the octupole Hamiltonian equations (\ref{eqfinal})). In the following, this feature is analyzed in more detail.

To understand this particular behavior, we decide to realize an analytical study of the frequencies of the system, similar to the one of \cite{Lib08}. Using a 12th-order expansion of the perturbative potential in powers of the eccentricities and the inclinations, they have performed Lie transformations to introduce an action-angle formulation of the Hamiltonian and identify the analytical expressions of the four fundamental frequencies of the 3-D secular (non-resonant) three-body problem. This study has been realized in two reference frames: a general one and the Laplace plane reference frame. Our aim in the present section is to wonder whether the dynamics around $35^\circ$ pointed out in Figures \ref{35} and \ref{lgterme}, corresponds to a commensurability between the fundamental frequencies. For the sake of completness, their analytical study is briefly described here.

The Hamiltonian function expanded in powers of the eccentricities and the inclinations and averaged over the mean anomalies $M_i$  reads: 
\begin{equation}\label{hamilt1}
\mathcal{K}= -\frac{Gm_1m_2}{a_2} \displaystyle{\sum_{k,j_{1},j_{2},i_l,l \in \underline{4}}}
B_{i_{l}}^{k,j_{1},j_{2}} E_1^{|j_1|+2i_1}E_2^{|j_2|+2i_2} S_1^{|k+j_{1}|+2i_3}S_2^{|k+j_2|+2i_4}\cos\Phi,
\end{equation}
with $\Phi=[j_1p_1-j_2p_2-(k+j_{1})q_{1}+(k+j_{2})q_{2}]$, $E_i=\sqrt{2P_i/L_i}$ and $S_i=\sqrt{2Q_i/L_i}$. The canonical variables in formula (\ref{hamilt1}) are the classical modified Delaunay's elements: 
\begin{equation}
\begin{array}{ll}
\lambda_i=\mbox{\rm mean longitude}&\quad L_i=m_i \sqrt{Gm_0a_i} \\ 
p_i=\mbox{\rm - longitude of the pericenter}&\quad P_i=L_i \left[ 1-\sqrt{1-{e_i}^2}
\right] \\
q_i=\mbox{\rm - longitude of the node} &\quad Q_{i}=L_{i} \sqrt{1-{e_{1}^{2}}} \left[1-\cos{i_{i}}\right]. \\
\end{array}
\end{equation}
The indices ($k,i_l,l \in \underline{4}$) are positive integers. The coefficients $B_{i_{l}}^{k,j_{1},j_{2}}$ depend only on the ratio $a_1/a_2$ of the semi-major axes. The secular Hamiltonian is a four degree of freedom problem. Let us note that it only depends on three angles, as 
\begin{equation}\label{3angles}
\Phi=j_{1}(p_{1}-q_{1})-j_{2}(p_{2}-q_{2})-k(q_{1}-q_{2}).
\end{equation}
As shown in \cite{Lib07}, the numerical convergence of the secular series (\ref{hamilt1}) is very good for a large set of parameters, even for moderate values of the eccentricities and the inclinations. The development is limited to order 12 in the eccentricities and the inclinations, which means that are kept in the Hamiltonian the terms for which the sum of the exponents of $E_1$, $E_2$, $S_1$ and $S_2$ is lower or equal to 12. 

In order to obtain the analytic expressions of the four fundamental frequencies, they have used a Lie transform perturbation scheme (Deprit 1969) to average the Hamiltonian (\ref{hamilt1}) over the secular variables $p_i'$ and $q_i'$ (i.e. the secular variables after a ``reducing rotation'' (\citealt{Hen88})). After a second Lie transform on the combination $\bar{p}'_{1}+\bar{p}'_{2}+\bar{q}'_{1}-3\bar{q}'_{2}$, they get the following action-angle formulation of the Hamiltonian -- we refer to \citet{Lib08} for more detail--:
\begin{equation}\label{hamilt2}
{\bar{{\mathcal K}'}}  =  \displaystyle{\sum_{l_{1}+l_{2}+l_{3}\le12}} {C}_{l_{1},l_{2},l_{3}} \; {\bar {E'_1}}^{2l_1} {\bar{E'_2}}^{2l_2} {\bar{S'_1}}^{2l_3}. 
\end{equation}
The associated Hamiltonian equations lead to the expression of the four frequencies:
\begin{equation}\label{freqqq}
\begin{array}{l}
\dot{\bar{p_{1}}}'=-\frac{(1-\mu)}{ \sqrt{\alpha}} \displaystyle{\sum_{l_i,i\in \underline{3}}} 2l_{1} {C}_{l_{1},l_{2},l_{3}} \bar{E'_1}^{2(l_1-1)} \bar{E'_2}^{2l_2}  \bar{S'_1}^{2l_3}  \\
\dot{\bar{p_{2}}}'=-\mu \displaystyle{\sum_{l_i,i\in \underline{3}}} 2l_{2}{C}_{l_{1},l_{2},l_{3}}\bar{E'_1}^{2l_1} \bar{E'_2}^{2(l_2-1)} \bar{S'_1}^{2l_3}  \\
\dot{\bar{q_{1}}}'=-\frac{(1-\mu)}{ \sqrt{\alpha}} \displaystyle{\sum_{l_i,i\in \underline{3}}} 2l_{3} {C}_{l_{1},l_{2},l_{3}}  \bar{E'_1}^{2l_1}\bar{E'_2}^{2l_2}  \bar{S'_1}^{2(l_3-1)} \\
\dot{\bar{q_{2}}}'= 0,
\end{array} 
\end{equation}
$\mu$ being the mass ratio $m_1/(m_1+m_2)$. 
The unit of frequency is the keplerian frequency $n_2=\sqrt{Gm_0/a_2^3}$ of the mass $m_2$ multiplied by the mass ratio $(m_1+m_2)\,/ \,m_0$. Let us note that, in the Laplace plane reference frame, the long-term evolution of the orbital elements can be described by only two frequencies and their linear combinations:  $f_1=-\dot{\bar{p}}'_{1}+\dot{\bar{q}}'_{1}$ and $f_2=-\dot{\bar{p}}'_{2}+\dot{\bar{q}}'_{1}$.

\begin{table}    
  \caption{Long-term evolution of a system with $i=30^\circ$, obtained by decompositions of frequencies on the data sets of the octupole approximation. Periods are expressed in years. Initial parameters of the system are $e_1=10^{-6}$, $e_2=0.3$, $\alpha=0.05$ and $m_1/m_2=10^{-4}$.}
    \begin{center}
      \begin{tabular}{r|cccc|ll}
        \hline
        Periods & $e$ & $\omega_1$ & $\omega_2$ & $\Delta\omega$ & &  \\
        \hline
        301\,753 & $c_1$ & $c_2$ & $c_4$ & ${\bf c_1}, c_2$ & $-\dot{\bar{p}}'_{1}+\dot{\bar{p}}'_{2}$ & $f_1-f_2$ \\
        51\,826 & $c_2$ & & $c_3$ & & $-\dot{\bar{p}}'_{1}-\dot{\bar{p}}'_{2}+2\dot{\bar{q}}'_{1}$ & $f_1+f_2$\\
        44\,229 & $c_3$ & $c_5$ & $c_2$ & $c_5$ & $-2\dot{\bar{p}}'_{1}+2\dot{\bar{q}}'_{1}$ & $2f_1$\\
        150\,876 & $c_4$ & $c_3$ & & $c_3$ & $-2\dot{\bar{p}}'_{1}+2\dot{\bar{p}}'_{2}$ & $2f_1-2f_2$\\
        38\,575 & $c_5$ & & & & $-3\dot{\bar{p}}'_{1}+\dot{\bar{p}}'_{2}+2\dot{\bar{q}}'_{1}$ & $3f_1-f_2$ \\
        88\,459 & & ${\bf c_1}$ & & & $-\dot{\bar{p}}'_{1}+\dot{\bar{q}}'_{1}$ & $f_1$ \\
        100\,584 & & $c_4$ & & $c_4$ & $-3\dot{\bar{p}}'_{1}+3\dot{\bar{p}}'_{2}$ & $3f_1-3f_2$ \\
        125\,146 & & & ${\bf c_1}$ & & $-\dot{\bar{p}}'_{2}+\dot{\bar{q}}'_{1}$ & $f_2$ \\
        62\,573 & & & $c_5$ & & $-2\dot{\bar{p}}'_{2}+2\dot{\bar{q}}'_{1}$ & $2f_2$\\
        \hline
      \end{tabular}
    \end{center}
  \label{table30}
\end{table}

In the following, we study the evolution of these frequencies with increasing values of the mutual inclination between the orbital planes. By resorting to a frequency analysis (\citealt{Las93}) on the data sets obtained with the octupole approximation, Table~\ref{table30} identifies the main combinations of the fundamental frequencies common to the evolutions of the orbital elements for $i=30^\circ$. The frequencies are listed by decreasing amplitude of the trigonometric term and noted $c_1$ (highest amplitude) to $c_5$. Bold type $c_1$ corresponds to the precession rate of an angular variable in circulation. Last columns display the identifications of the different combinations in terms of the fundamental frequencies ($\dot{\bar{p}}'_{1}$, $\dot{\bar{p}}'_{2}$ and $\dot{\bar{q}}'_{1}$) and the two frequencies $f_1$ and $f_2$ respectively. 

As can be observed in Table \ref{table30}, the two frequencies, $f_1=0.01184$ and $f_2=0.00837$ (values calculated from equation (\ref{freqqq}) of the analytical 12th-order expansion, in their unit of frequency), correspond to the precession rates of the arguments of the pericenter $\omega_1$ and $\omega_2$ respectively. Last column shows that all the frequencies of the orbital evolutions are linear combinations of these two frequencies. In particular, the main frequency of the eccentricities is the precession rate of $\Delta \omega$ and corresponds to $f_1-f_2$. Let us note that the analytical frequencies given by equation (\ref{freqqq}) are very close to the ones identified by the frequency analysis on the octupole approximation: $f_{1oct}=0.01192$ and $f_{2oct}=0.00839$. This small shift in the periods (less than $10^3$ years) is due to the limitations of both approximations with respect to the semi-major axes ratio or the eccentricities and inclinations.

\begin{figure}
\centering{
\rotatebox{270}{\includegraphics[height=8.5cm]{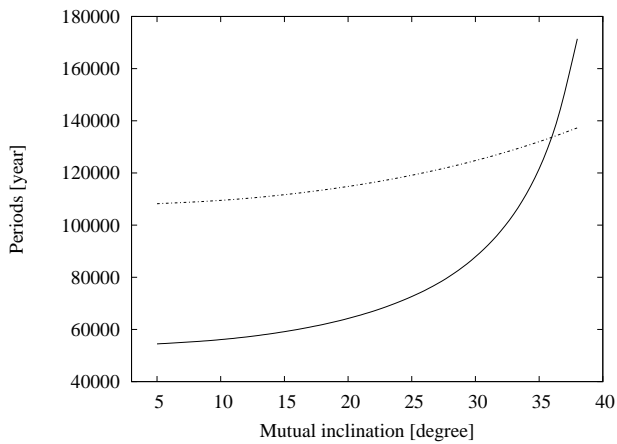}}
}
\caption{Evolution of the frequencies $f_1$ (solid line) and $f_2$ (dot-dashed line) for increasing values of the mutual inclination}
\label{freqfig}
\end{figure}

To analyze the dynamics around a mutual inclination of $35^\circ$ in more detail, we examine the evolution of the two frequencies for increasing mutual inclination values. Figure \ref{freqfig} shows the evolution of the periods associated to $f_1$ and $f_2$ for mutual inclination up to $38^\circ$. The two curves intersect when $i=\sim36^\circ$, namely $35.55^\circ$ in the octupole formulation and $35.9^\circ$ in the eccentricities and inclinations development.

\begin{table}
  \caption{Same as Table \ref{table30} for $i=35.55^\circ$}
    \begin{center}
      \begin{tabular}{r|cccc|l}
        \hline
        Periods & $e$ & $\omega_1$ & $\omega_2$ & $\Delta\omega$ &   \\
        \hline
        66\,491 & $c_1$ & $c_2$ & $c_3$ & $c_1$ & 2f  \\
        4\,881\,857 & $c_2$ & $c_4$ & $c_2$ & $c_3$ & g  \\
        67\,409 & $c_3$ & & $c_4$ & & 2f-g  \\
        33\,245 & $c_4$ & $c_3$ & & $c_2$ & 3f  \\
        65\,597 & $c_5$ & $c_5$ & & $c_4$ & 2f+g  \\
        132\,982 & & ${\bf c_1}$ & ${\bf c_1}$ & & f  \\
        2\,440\,933 & & & $c_5$ & & 2g  \\
        33\,020 & & & & $c_5$ & 4f+g  \\
        \hline
      \end{tabular}
    \end{center}
  \label{table35}
\end{table}

\begin{figure}
\centering{
\rotatebox{270}{\includegraphics[width=11cm]{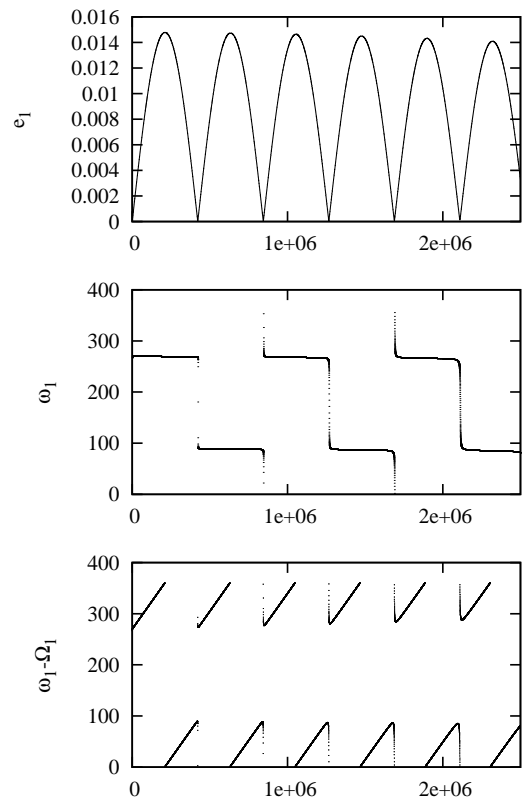}}
}
\caption{Long-term evolution of a system with mutual inclination of $35.55^\circ$ in a frame where the reference plane is the plane of the giant planet. Note that the resonant angle is $\omega_1-\Omega_1$.}
\label{planfunk}
\end{figure}

As a result, the coupling between the frequencies of the orbital elements is different for this particular value of the mutual inclination, as shown in Table \ref{table35} (frequency analysis on the octupole approximation). Indeed, all the frequencies are combinations of $f=f_1=f_2$ and a very small frequency $g$. The change of dynamics induced by the commensurability $f_1=f_2$, is obvious when looking at the $\Delta \omega$'s evolution in Figure \ref{lgterme}: the angle is in libration for the very low eccentric orbit at $i=35.55^\circ$. As it corresponds to a behavior modification of an angle, this particular dynamics can thus be regarded as a secular resonance.

All this study is realized in the Laplace plane reference frame. If we consider the orbital plane of the giant planet as reference plane ($i_2=0$), the evolution of the eccentricities is similar and the resonant angle becomes $\omega_1-\Omega_1$, as illustrated in Figure \ref{planfunk}.

\section{Conclusion}

In the present work, we focused on the study of the 3-D elliptic three-body problem of a small mass under the attraction of an outer giant body. The influence of an eccentric orbit of the perturber on the dynamics of a small inclined inner body has, to our knowledge, not yet been investigated in the literature. Particular attention has been given to a region around $35^\circ$ of mutual inclination detected numerically by \citet{Fun11}.  

Our analytical study relies on the octupole expansion, which is a compact formulation of the Hamiltonian suitable for hierarchical planetary systems. Short-period averaging and node reduction (by adoption of the Laplace plane reference frame) enable us to reduce the problem to two degrees of freedom. The four-dimensional dynamics is analyzed through representative planes which identify the main equilibria of the problem. It has been shown that an inner body on quasi-circular orbit behaves secularly in a similar way as in the circular three-body problem: its eccentricity variations are very limited for mutual inclination between the orbital planes smaller than $\sim 40^\circ$, while they become large and chaotic for higher mutual inclination.

As shown by \citet{Fun11}, there exists a dynamical region around $35^\circ$ of mutual inclination consisting of long-time stable and particularly low eccentric orbits of the small body. Using a 12th-order Hamiltonian expansion in eccentricities and inclinations, in particular its action-angle formulation obtained by Lie transforms in \citet{Lib08}, we have shown that this region corresponds to a commensurability of the two frequencies that are the precession rates of the arguments of the pericenter $\omega_1$ and $\omega_2$. It explains the change of dynamics of the angle $\Delta \omega$ which starts to evolve in libration. This particular dynamics can thus be regarded as a secular resonance. The same analysis can be realized with the orbital plane of the giant planet as the reference plane (i.e. no adoption of the Laplace plane) to identify $\omega_1-\Omega_1$ as the resonant angle of this reference frame. 

This study also applies to binary star systems, where a planet is revolving around one of the two stars (inner problem), since the mass ratio between the bodies is of the same order ($\mu \sim 10^{-4}$).

The region around $35^\circ$ could belong to the habitable zone of the system and be of particular interest for the research of life in extrasolar systems, as it consists of stable orbits with limited eccentricity variation of the planet, which means a constant distance between the planet and the host star.

\section*{Acknowledgement}
The work of A.-S. Libert is supported by an F.R.S.-FNRS Postdoctoral Research Fellowship. A.-S. Libert warmly thanks the team of Vienna for fruitful discussions. Numerical simulations were made on the local computing resources (Cluster URBM-SYSDYN) at the University of Namur (FUNDP, Belgium).


\begin{thebibliography}{}


\bibitem[\protect\citeauthoryear{Delsate et al.}{2010}]{Del10}
Delsate, N., Robutel, P., Lemaitre, A., Carletti, T., 2010, CeMDA, 108, 275

\bibitem[\protect\citeauthoryear{Deprit}{1969}]{Dep69}
Deprit, A., 1969, Celestial Mech., 1, 12

\bibitem[\protect\citeauthoryear{Fabrycky \& Tremaine}{2007}]{Fab07}
Fabrycky, D., Tremaine, S., 2007, ApJ, 669, 1298
 
\bibitem[\protect\citeauthoryear{Farago \& Laskar}{2010}]{Far10}
Farago, F., Laskar, J., 2010, MNRAS, 401, 1189

\bibitem[\protect\citeauthoryear{Ferrer \& Os\'acar}{1994}]{Fer94}
Ferrer, S., Os\'acar, C., 1994, CeMDA, 58, 245 

\bibitem[\protect\citeauthoryear{Ford et al.}{2000}]{For00}
Ford, E.B., Kozinski, B., Rasio, F.A., 2000, ApJ, 535, 385

\bibitem[\protect\citeauthoryear{Funk et al.}{2011}]{Fun11}
Funk, B., Libert, A.-S., S\"uli, \'A., Pilat-Lohinger, E., 2011, A\&A, 526, id.A98

\bibitem[\protect\citeauthoryear{Harrington}{1969}]{Har69}
Harrington R.S., 1969, Celest. Mech., 1, 200

\bibitem[\protect\citeauthoryear{Henrard}{1988}]{Hen88}
Henrard, J., 1988, Celest. Mech., 45, 327

\bibitem[\protect\citeauthoryear{Innanen et al.}{1997}]{Inn97}
Innanen, K. A., Zheng, J. Q., Mikkola, S., Valtonen, M. J., 1997, AJ, 113, 1915 

\bibitem[\protect\citeauthoryear{Jacobi}{1842}]{Jac42}
Jacobi C.G.J., 1842, Astronomische Nachrichten, 20, 81

\bibitem[\protect\citeauthoryear{Kozai}{1962}]{Koz62}
Kozai, Y., 1962, ApJ, 67, 591

\bibitem[\protect\citeauthoryear{Laskar}{1993}]{Las93}
Laskar, J., 1993, Phys. D, 67, 257

\bibitem[\protect\citeauthoryear{Laskar}{1997}]{Las97}
Laskar, J., 1997, A\&A, 317, L75

\bibitem[\protect\citeauthoryear{Laskar \& Robutel}{1995}]{Las95}
Laskar, J., Robutel, 1995, CeMDA, 62, 193

\bibitem[\protect\citeauthoryear{Laskar \& Bou\'e}{2010}]{Las10}
Laskar, J., Bou\'e, G., 2010, A\&A, 522, id.A60

\bibitem[\protect\citeauthoryear{Lee \& Peale}{2003}]{Lee03}
Lee, M.H., Peale, S.J., 2003, ApJ, 592, 1201

\bibitem[\protect\citeauthoryear{Libert \& Henrard}{2007}]{Lib07}
Libert, A.-S., Henrard, J., 2007, Icarus, 191, 469

\bibitem[\protect\citeauthoryear{Libert \& Henrard}{2008}]{Lib08}
Libert, A.-S., Henrard, J., 2008, CeMDA, 100, 209

\bibitem[\protect\citeauthoryear{Libert \& Tsiganis}{2009}]{Lib09}
Libert, A.-S., Tsiganis, K., 2009, A\&A, 493, 677

\bibitem[\protect\citeauthoryear{Lidov}{1962}]{Lid62}
Lidov, M.L., 1962, Planetary and Space Science, 9, 719

\bibitem[\protect\citeauthoryear{Lidov \& Ziglin}{1976}]{Lid76}
Lidov, M.L., Ziglin, S.L., 1976, Celest. Mech., 13, 471

\bibitem[\protect\citeauthoryear{Malige et al.}{2002}]{Mal02}
Malige, F., Robutel, P., Laskar, J., 2002, CeMDA, 84, 283

\bibitem[\protect\citeauthoryear{Michtchenko et al.}{2006}]{Mic06}
Michtchenko, T.A., Ferraz-Mello, S., Beaug\'e, C., 2006, Icarus, 181, 555

\bibitem[\protect\citeauthoryear{Migaszewski \& Go\'zdziewski}{2009}]{Mig09}
Migaszewski, C., Go\'zdziewski, K., 2009, MNRAS, 395, 1777

\bibitem[\protect\citeauthoryear{Migaszewski \& Go\'zdziewski}{2011}]{Mig11}
Migaszewski, C., Go\'zdziewski, K., 2011, MNRAS, 411, 565

\bibitem[\protect\citeauthoryear{Poincar\'e}{1896}]{Poi96}
Poincar\'e, H., 1896, C.R.A.S, 93, 1031

\bibitem[\protect\citeauthoryear{Prado}{2005}]{Pra05}
Prado, A.F.B.A, 2005, Braz. Soc. Mech. Sci. \& Eng., 27, 364
 
\bibitem[\protect\citeauthoryear{Russell \& Brinckerhoff}{2009}]{Rus09}
Russell, R.P., Brinckerhoff, A.T., 2009, J. Guid. Control Dyn., 32, 424

\bibitem[\protect\citeauthoryear{Thomas \& Morbidelli}{1996}]{Tho96}
Thomas, F., Morbidelli, A. 1996, CeMDA, 64, 209

\bibitem[\protect\citeauthoryear{Wu \& Murray}{2003}]{Wu03}
Wu, Y., Murray, N., 2003, ApJ, 589, Issue 1, 605


\end{thebibliography}

\bibliographystyle{}

\end{document}